\documentclass{aa} 

\usepackage{graphicx}  
\usepackage{siunitx}   
\usepackage{color}     
\usepackage{txfonts}   
\usepackage{longtable} 
\usepackage{mathrsfs}  

\usepackage[colorlinks=true, linkcolor=blue, citecolor=blue, urlcolor=blue]{hyperref}

\begin{document}

\title{Gaia and IRTF abundance of A-type main belt asteroids}

   \author{M. Delbo\inst{1,2}
          \and
          C. Avdellidou\inst{2,1}
          \and
          M. Galinier \inst{1,3}
          \and
          U. Bhat \inst{2}
          \and
          T. Dyer \inst{2}
          \and
          B. T. Bolin \inst{4}
          \and
          L. Galluccio \inst{1,5}
        }

   \institute{Université C\^ote d’Azur, CNRS–Lagrange, Observatoire de la C\^ote d’Azur, CS 34229 – F 06304 NICE Cedex 4, France\\
              \email{delbo@oca.eu}
         \and
             University of Leicester, School of Physics and Astronomy, University Road, LE1 7RH, Leicester, UK
         \and
             INAF-IAPS, Institute for Space Astrophysics and Planetology, Rome, Italy
          \and 
              Eureka Scientific, Oakland, CA 94602, USA
              \and
             Osservatorio Astronomico di Roma, Via Frascati 33, I-00078 Monte Porzio Catone, Italy \\
             }

   \date{Received August 11, 2025; accepted October 10, 2025}
 
 \abstract
{The so-called Missing Mantle Problem is a long-standing issue in planetary science. It states that olivine-rich asteroids should be abundant in the main belt, while this is observationally found not to be the case, by dedicated surveys. Conversely, olivine-rich asteroids appear to be more abundant among near-Earth asteroids than those surveys would suggest.}
{We aim to provide a revised estimate of the abundance of A-type (olivine-rich) asteroids in the main belt by combining taxonomic classifications from Gaia Data Release 3 reflectance spectra with ground-based near-infrared observations from NASA's IRTF.}
{We performed a principal component analysis (PCA) on Gaia Data Release 3 visible-light reflectance spectra to identify A-type candidates, and confirmed a subset of these using near-infrared spectroscopy from IRTF. We combined our observations with data from the literature to compute the A-type probability distribution as a function of the principal components of Gaia reflectance spectra. This probability distribution was then used to estimate the abundance of A-type asteroids in the main belt and its subpopulations as a function of heliocentric distance. We also examined the distribution of A-type asteroids among known collisional families.}
{We found that the abundance of A-types in the main belt is $(2.00 \pm 0.15)$\%, significantly higher than previous estimates for the same region. Our analysis also shows that some collisional families, such as those of Vesta and Flora, have above-average A-type fractions, whereas others, like Themis and Hygiea, exhibit negligible abundance.}
{Our results support the idea that olivine-rich material is more widespread than previously thought. In particular, the high A-type abundance in the Flora family is consistent with the hypothesis of a second differentiated parent body in the inner main belt, beyond Vesta. This work provides new observational constraints on the Missing Mantle Problem and the distribution of differentiated material in the asteroid main belt. In particular, our results deepen the compositional diversity observed in the inner main belt and have important implications for the understanding of early solar system differentiation processes.}


   \maketitle

\section{Introduction}

It is well understood that main belt asteroids are remnants of the planetesimals that escaped merging collisions to form the terrestrial planets \citep{morbidelliDidTerrestrialPlanets2025} and the dynamical ejection out of our solar system \citep[planetesimals are the so-called first $\sim$100~km sized bodies that accreted from the dust of our protoplanetary disc, as reviewed by][]{klahrFormationMainBelt2022}. 

Mutual collisions between asteroids created, continuously over time \citep{brozConstrainingCometaryFlux2013,spotoAsteroidFamilyAges2015}, families of fragments 
\citep{nesvornyIdentificationFamiliesAstIV2015,delboIdentificationPrimordialAsteroid2017,delboAncientPrimordialCollisional2019,ferrone2023,nesvornyCatalogProperOrbits2024} that we can still identify today. Family members dominate in number today's main belt asteroid population \citep{delboIdentificationPrimordialAsteroid2017,dermottCommonOriginFamily2018,ferrone2023} and are recognised as the main sources of meteorites \citep{greenwoodLinkingAsteroidsMeteorites2020,avdellidouAthorAsteroidFamily2022,brozSourceRegionsCarbonaceous2024,brozYoungAsteroidFamilies2024,marsset2024}. Hence, meteorites, together with samples returned from asteroids \citep{lauretta2024}, allow us to probe the materials that constituted the planetesimals \citep{bourdelledemicasCompositionInnerMainbelt2022}, as well as to infer information about the dynamical processes that sculpted our solar system \citep[see, e.g.,][]{avdellidouDatingSolarSystems2024}.

A fraction of meteorites have compositions consistent with origins from differentiated planetesimals \citep[][in particular their Table 2]{burbineMeteoriticParentBodies2002}. These are the bodies that underwent processes of internal heating strong enough to cause their materials, originally mixed together as undifferentiated (or chondritic), to separate into distinct layers, i.e., a core, a mantle, and a crust \citep[][and references therein]{neumannDifferentiationCoreFormation2012}.  This internal heating is thought to have originated from the decay of radioactive elements \citep[mainly $^{26}$Al, see e.g.][but also $^{60}$Fe]{ureyCosmicAbundances1955} that were spread over the protoplanetary disk \citep{macphersonDistributionAluminum26_1995} and therefore accreted into the early planetesimals.

In a fully differentiated object, the core is made of iron-nickel metal, the mantle is constituted of igneous silicates, such as olivine -- which is commonly found in mafic and ultramafic igneous rocks, and is composed of magnesium and iron silicate (Fe, Mg)$_2$SiO$_4$ -- and the crust is also composed of igneous rocks, such as basalt \citep{weissDifferentiatedPlanetesimalsParent2013}. The latter may have formed from rapid cooling of lava material exposed to space. 
Partially differentiated planetesimals would still preserve a crust of undifferentiated chondritic materials at various metamorphic grades \citep[see Figure 2 of][]{weissDifferentiatedPlanetesimalsParent2013}, overlying an olivine rich mantle and an iron core. The type of differentiation is primarily controlled by the onset and stop of the heating process \citep[see Figure 3 of][]{weissDifferentiatedPlanetesimalsParent2013}. The former is related to the epoch of planetesimal accretion and the latter is mostly controlled by the planetesimal size and thermal conductivity. The various degrees of differentiation explain the diversity of meteorites, which ranges from undifferentiated chondrites to metallic meteorites, including those with varying degrees of metamorphism.

Different types of meteorites have been successfully linked to different asteroid spectroscopic types by studying the features of their reflectance spectra, such as ordinary chondrites \citep[][and references therein]{brozYoungAsteroidFamilies2024,marsset2024}, basaltic meteorites \citep[i.e., crustal, see][]{moskovitzSpectroscopicallyUniqueMainBelt2008,solontoiAVASTSurvey04102012,leithCompositionalDiversityNonVesta2017}, enstatite meteorites \citep{avdellidouAthorAsteroidFamily2022}, carbonaceous chondrites \citep{brozSourceRegionsCarbonaceous2024} and metallic meteorites \citep{Avdellidou2025MNRAS.539.3534A} \citep[the latter population also from analysis of spectroscopic and thermal infrared data, see][]{harrisHowFindMetalrich2014}.

However, it has been claimed that there is a shortage of olivine-dominated mantle material -- a tracer of differentiation -- in the main belt \citep{demeoOlivinedominatedAtypeAsteroids2019}. This conundrum is known as the ``Missing Mantle Problem'', or the ``Great Dunite Shortage'', which is a long-standing issue in planetary science \citep{Chapman1986MmSAI..57..103C,Bell1989aste.conf..921B}, stating that olivine-rich asteroids should be abundant in the main belt, while this is observationally found not to be the case \citep[see also, e.g.,][]{demeoOlivinedominatedAtypeAsteroids2019}. 

Olivine-rich asteroids are typically identified by having spectroscopic A-type class in the Bus and Binzel \citep{busPhaseIISmall2002} and/or in the Bus-DeMeo \citep{demeoExtensionBusAsteroid2009} taxonomies. The latter is more reliable due to its extension in the near-infrared (NIR) that is very diagnostic for this class. Namely, A-types have very high reflectance in the NIR with no or very weak 2~\SI{}{\micro\meter} absorption band in addition to be also characterised by reference spectra with very red slopes at wavelengths $\lesssim$0.7~\SI{}{\micro\meter}, and a strong absorption feature centred slightly longwards of 1~\SI{}{\micro\meter} \citep[e.g.,][]{busPhaseIISmall2002,demeoExtensionBusAsteroid2009,sanchezOlivinedominatedAsteroidsMineralogy2014,demeoOlivinedominatedAtypeAsteroids2019}. Moreover, A-types have a moderate to high geometric visible albedo values ($p_\text{V}$=0.26$^{+0.11}_{-0.08}$), as we show in section section \ref{S:albedo}. A-types show the closest spectroscopic similarities to brachinite and pallasite meteorites \citep{burbineSmallMainBeltAsteroid2002}. Since these meteorites originated in the mantle and in the core-mantle boundary of differentiated planetesimals \citep{benedixIronStonyIronMeteorites2014}, it is reasonable to assume that A-type asteroids could represent fragments from these layers of a differentiated planetesimal.

The most recent survey of A-types in the main belt \citep{demeoOlivinedominatedAtypeAsteroids2019} made use of the Sloan Digital Sky Survey (SDSS) Moving Object Catalog visible (VIS) photometric data to select a number of candidate A-type asteroids; then NIR spectroscopy of the candidates by means of SpeX on the NASA Infrared Telescope Facility (IRTF) and the Folded-port InfraRed Echellette (FIRE) on the Magellan Telescope was performed to confirm their A-type classes. By multiplying the confirmation rate and the number of candidates compared to the total asteroid population, this survey found that A-types represent only a 0.16\% of the main belt asteroids with sizes $\geq$2~km. These results lead to conclude that the Missing Mantle Problem is a fact and that asteroid differentiation was not as widespread as previously thought \citep{demeoOlivinedominatedAtypeAsteroids2019}. Moreover, it was found that A-types are evenly distributed throughout the main belt, even detected at the distance of the Cybele region, and have no statistically significant concentration in any asteroid collisional family.  

However, \cite{galinierDiscoveryFirstOlivinedominated2024} discovered that the outer main belt family (36256) 1999 XT17 presents a prominence of A-type objects, thus ruling out the hypothesis of \cite{demeoOlivinedominatedAtypeAsteroids2019} that A-types have no statistically significant concentration in any collisional family. In addition to (36256) 1999 XT17, \cite{galinierDiscoveryFirstOlivinedominated2024} indicate other collisional families potentially having abundance of A-types. 

\cite{marsset2022} combined NIR spectra of near-Earth objects (NEOs) collected as part of the MIT-Hawaii NEO Spectroscopic Survey with data retrieved from \cite{binzel2019} to derive the intrinsic compositional distribution of the overall NEO population. Their bias-corrected distribution shows that the A-types account for the 0.1-0.5\% of the sample coming from each different NEO source region.

Moreover, \cite{sergeyevCompositionalPropertiesPlanetcrossing2023} used spectrophotometric data from the SDSS and SkyMapper surveys and reported that A-type asteroids constitute 2.5$\pm$0.2\% of the near-Earth asteroid (NEA) population. This fraction is more than 15 times higher than the A-type abundance derived by \cite{demeoOlivinedominatedAtypeAsteroids2019} for the main asteroid belt. \cite{sergeyevCompositionalPropertiesPlanetcrossing2023} also find that A-types show a higher number density in the region between the orbit of Mars and Jupiter's 4:1 mean motion resonance, which is predominantly occupied by the Hungaria asteroid population. These results suggested a potential link between Hungaria asteroids and A-type NEAs.

Similar findings were reported by \cite{devogeleVisibleSpectroscopyMission2019}. They estimated that A-types account for 3.8$\pm$1.3\% of the NEA population. They identified an excess of A-types at semimajor axes around 1.6~au, which corresponds to the 1:2 mean motion resonance with Earth. This resonance likely facilitates the transfer of Hungaria family asteroids into the NEA population. Further supporting these observations, \cite{lucasHungariaAsteroidRegion2019} conducted a spectroscopic survey of the Hungaria region and found an A-type fraction of 1.5\%. 
Furthermore, \cite{popescuTaxonomicClassificationAsteroids2018} identified an abundance of 5.4\% of A-types in a sample of 147 NEAs with diameters less than 300~m, which were surveyed in the visible wavelength range. Similarly, \citet{Perna2018P&SS..157...82P} found an abundance of A-types of 5.48\% among 146 NEAs based on ground-based observations also in the visible range.

However, we should treat cautiously A-type classification based only on the VIS spectrum: \cite{demeoOlivinedominatedAtypeAsteroids2019} showed that their typical A-type confirmation rate using NIR spectroscopy is about 1/3 of the potential A-types identified from VIS spectrophotometry (SDSS). In addition, care should be used when comparing directly the main belt surveys with the NEO ones, because they are potentially affected by different observing biases. For instance, in the case of NEOs, by focusing on targets on specific orbits (e.g. low-$\Delta$v) a bias in favour of objects coming from specific NEO source regions in the main belt may be applied. This can result in samples that are not representative of the overall main belt population. Moreover, NEO and main-belt surveys probe different size ranges, which could affect the inferred A-type fractions.

All the results presented above indicate that the Missing Mantle Problem might need a review in the light of new astronomical data. Gaia Data Release 3 (Gaia DR3) provided to the community an unprecedented dataset of more than 60,000 asteroid visible reflectance spectra \citep{gaiacollaborationGaiaDataRelease2023}. Such a dataset can be used to perform new searches for the missing olivine in the main belt \citep{galinierDiscoveryFirstOlivinedominated2024}. Namely, we used Gaia DR3 data to inform target selection for a spectroscopic survey of main belt asteroids in the NIR (using NASA's IRTF) that we combined with existing literature data to test Gaia DR3 predictions for A-type asteroids. 

In Sect. 2 we present the Gaia DR3 dataset, in Sect. 3 the target selection for the NIR observations, in Sect. 4 the analysis and results which we discuss in Sect. 5.

\section{Data and target selection}
We made use of VIS reflectance spectra that were derived from asteroid spectroscopic observations obtained by the Gaia space mission of the European Space Agency (ESA) between 5 August 2014 and 28 May 2017 and were published as part of the Gaia DR3 
\citep{gaiacollaborationGaiaDataRelease2023,tangaGaiaDataRelease2023}. It contains 60,518 solar system small bodies, mostly main belt asteroids. To each asteroid is associated a unique mean reflectance spectrum obtained by averaging several epoch spectra, which span the VIS wavelength range between 0.374 and 1.034~\SI{}{\micro\meter}. All mean reflectance spectra are expressed by the same 16 discrete wavelength bands. A `reflectance\_spectrum\_flag' (RSF) is associated with each band, signalling whether the band reflectance value and its uncertainty are validated (RSF=0), suspected to be of poor quality (RSF=1), or not good (RSF=2). There is evidence that Gaia DR3 reflectance spectra overestimate the reflectance values in the reddest bands \citep{gaiacollaborationGaiaDataRelease2023,galinierGaiaSearchEarlyformed2023,galinierDiscoveryFirstOlivinedominated2024}, which however, appears not to cause problems in the identification of A-types \citep{galinierDiscoveryFirstOlivinedominated2024}. The two bluest bands are also often affected by large uncertainties and have often RSF>0 \citep{gaiacollaborationGaiaDataRelease2023,delboGaiaViewPrimitive2023}. 
\cite{tinaut-ruanoAsteroidsReflectanceGaia2023} showed that the set of solar analogues adopted by \cite{gaiacollaborationGaiaDataRelease2023} yields asteroid reflectances with systematically redder spectral slopes at wavelengths shorter than 0.55~\SI{}{\micro\meter} compared to using Hyades 64 alone, the latter being considered the most Sun-like in the near-ultraviolet. They proposed a correction for reflectance bands shorter than 0.55~$\SI{}{\micro\meter}$. We do not apply this correction here, because it does not affect A-type identification, as this relies primarily on wavelengths longer than 0.55~$\SI{}{\micro\meter}$.

Our aim was to identify A-types in the Gaia DR3 dataset. One could do so by firstly performing a classification of the Gaia DR3 spectra and then identifying the objects that belong to the A-class. However, because of the systematic effects described above, we preferred to follow another established method: we carried out a Principal Component Analysis (PCA) of the Gaia DR3 reflectance spectra, which is a linear transformation from the reflectance space to another coordinate space where the greatest variance lies along the first axis (principal component), the second greater variance along the second axis, and so on, effectively capturing the most important patterns in the data. Clusters of points in the PCA space are expected to correspond to taxonomic classes (Appendix~D). This is similar to the approach used by \cite{demeoOlivinedominatedAtypeAsteroids2019} for their identification of potential A-types in the SDSS, and by \cite{choiTaxonomicClassificationAsteroids2023} for their taxonomic classification using the KMTNet multi-band photometry. 

Before presenting our methods, we describe the data selection procedure: (i) we applied the PCA only on Gaia DR3 reflectance spectra with S/N~$>$50 \citep[where the S/N is calculated as in][]{gaiacollaborationGaiaDataRelease2023} to ensure that the PCA transformation was not biased by noisier data. This resulted defining the PCA transformation on 9,332 objects, which we then applied on the whole DR3 dataset, regardless of the S/N. We also tested lower S/N thresholds, e.g., 30 and 20 -- corresponding to 20,866 and 36,551 asteroids, respectively -- and found that the results did not change appreciably. We excluded the two reddest and two bluest wavelength bands in the PCA fit, and also discarded reflectance bands with RSF$>$~0. We retained five principal components in total, but found that the first and second components, hereafter referred to as PC1 and PC2, respectively, carry most of variance of the data. Namely, we found that the variance of the PC1--5 components are, respectively, 0.057, 0.017, 0.007, 0.004, 0.004. As expected, the position of asteroids in the PC1--PC2 plot is diagnostic of their compositional spectral class (Fig.~\ref{F:pcaClasses}). 

\begin{figure}
   \includegraphics[width=\columnwidth]{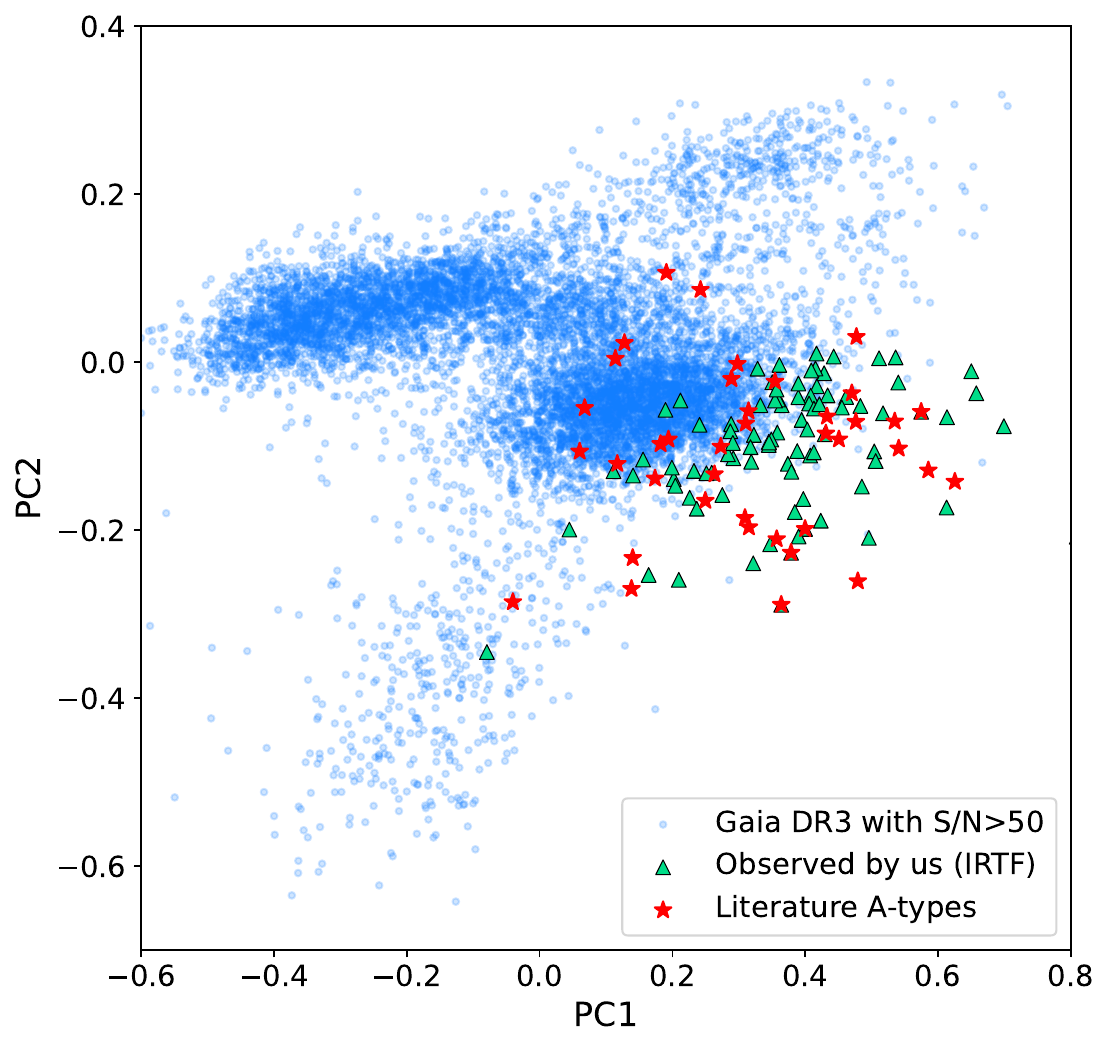}
   \caption{PC1 vs. PC2 plot of main belt asteroid reflectance spectra from the Gaia DR3 having S/N$>$50 (blue circles) \citep[see][for details of S/N estimation]{gaiacollaborationGaiaDataRelease2023}. 
   The red stars indicate the A-types that we selected from the literature following the method explained in the text and have also a DR3 reflectance spectrum. The green triangles show the asteroids that we observed with the IRTF.}
   \label{F:PCA50}
\end{figure}

Next, we needed a ground-truth sample of A-types with Gaia DR3 data to locate them in the PCA space. NIR spectroscopy is the most robust A-type identification method because it is sensitive to the distinguishing features of this class, as explained in the introduction. 
Hence, we retrieved from the literature all main belt asteroids that were classified A-type from both VIS and NIR spectroscopy (VIS+NIR) and have a corresponding Gaia DR3 reflectance spectrum. Our search resulted in 44 asteroids (Table~\ref{T:literatureAtypes}), which is a small set compared to the 1,755 asteroids that have spectral classes derived from VIS and NIR spectroscopy (A-type and non A-type). The PCA was calculated on Gaia DR3 having S/N$>$50, but asteroids were retrieved from literature or selected for NIR observations regardless of the S/N of their Gaia DR3 reflectance spectrum. The PC1 and PC2 values of A-types revealed that they are preferentially located at large PC1 and negative PC2 values (Fig.~\ref{F:PCA50}), adjacent to the S-complex cluster (Appendix~D). The latter is centred at PC1$\sim$~0.17, and PC2$\sim$~-0.06, as it can be seen in the S-type plot of Fig.~\ref{F:pcaClasses}. 

\section{Working hypothesis and methods}
\label{S:workingHypothesis}
All the above considerations lead us to formulate the hypothesis that we could identify A-types from their position in the PC1--PC2 plane. The values along the PC1 coordinate are strongly correlated with the overall spectral slope, while the PC2 coordinate is  correlated with the depth of the silicate absorption band centred near 1~\SI{}{\micro\meter}. This correlation is not new, as the position of asteroids in the two-dimensional space defined by spectral slope and Z–I magnitudes, the latter being related to the depth of the silicate band, has previously been used to perform spectral classification \citep{demeoTaxonomicDistributionAsteroids2013} of asteroids with SDSS spectrophotometry. However, while those authors applied fixed boundaries in the plane slope--Z-I for their classifications, here we adopted a different approach, namely, a method based on calculation of the probability that an asteroid is an A-type (or not) as function of its values of the principal components.  

In order to increase the number statistics from the literature (44 A-types), we performed NIR spectroscopy (section \ref{S:NIR@IRTF}) of 88 asteroids that have Gaia DR3 reflectances with PC1 and PC2 values preferentially located in the area where most of the known A-types plot (Fig.~\ref{F:PCA50}), regardless of Gaia DR3 reflectance S/N. We used NIR spectroscopy for the reason explained in the previous subsection. 
Next, we combined NIR reflectance spectra with VIS Gaia DR3 \citep[as described by][]{avdellidouAthorAsteroidFamily2022} and classified the combined spectra in the Bus-DeMeo taxonomy \citep{demeoExtensionBusAsteroid2009}, in order to follow previous studies and we assessed which are A-types and which are not (Sect.~\ref{S:classification}). Eventually, we combined our survey with literature data and defined two discrete sets of asteroids (A-type and non A-type) based on which we calculated the probability of an asteroid to be an A-type, on the basis of its PC1 and PC2 values derived from its Gaia DR3 reflectance spectrum (Appendix \ref{A:appendix}). In the following we describe in details our methods.

\subsection{Near-infrared observations}
\label{S:NIR@IRTF}

Between August 2022 and October 2024, we obtained NIR spectra for 88 asteroids whose PC1 and PC2 values lie mainly in the region where the probability of finding A-types is high (see Sect.~\ref{S:workingHypothesis}). 
Observations were conducted using the SpeX spectrograph \citep{raynerSpeXMediumResolution2003} and MORIS \citep{gulbisFirstResultsMIT2011}, the latter operating as an auto-guiding instrument, at the IRTF. SpeX was used in PRISM mode with a 0.8\arcsec$\times$15\arcsec slit, covering wavelengths from 0.7 to 2.5~\SI{}{\micro\meter} at a spectral resolution of $\sim$200 in a single configuration. Observational details are given in Table~\ref{T:ourIRTFObs}.

Spectra were reduced using Spextool (v4.1), an IDL-based spectral reduction tool \citep{cushingSpextoolSpectralExtraction2004}. Following standard procedures \citep{reddyComposition298Baptistina2009,avdellidouAthorAsteroidFamily2022}, we calculated the reflectance $R(\lambda)$ of each asteroid as a function of wavelength $\lambda$ using Eq.~\ref{E:spec}.

\begin{equation}
R(\lambda) = \frac{A(\lambda)}{S_L(\lambda)} \times 
	              Poly \left (\frac{S_L(\lambda)}{S_T(\lambda)} \right),
	              \label{E:spec}
\end{equation}
where $A(\lambda)$, $S_L(\lambda)$, $S_T(\lambda)$ are the wavelength-calibrated raw spectra respectively of the asteroid, a local G2-type star observed within $\sim$300" of the asteroid, and a well studied solar analogue (SA) star that was observed at similar airmass when possible, respectively. The function $Poly()$ represents a polynomial fit of the stars ratio, excluding the regions affected by the telluric water vapour absorption (1.3 $< \lambda <$ 1.5, 1.78 $ < \lambda <$ 2.1, and $\lambda > $ \SI{2.4}{\micro\meter}).
Asteroid spectra were shifted to sub-pixel accuracy to align with the calibration star spectra. In general, the local star ensures accurate removal of the telluric features, but may require the slope correction 
$Poly \left (\frac{S_L(\lambda)}{S_T(\lambda)} \right)$ 
due to difference between the local star's spectrum and that of the Sun. When the local star was not observed, the correction of the telluric features was performed directly by dividing the asteroid spectrum by that of the trusted solar analogue i.e: $R(\lambda) = \frac{A(\lambda)}{S_T(\lambda)}$. Table~\ref{T:ourIRTFObs} contains the relevant observing information.

\subsection{Classification}
\label{S:classification}
The NIR spectra of 88 main-belt asteroids were combined with Gaia DR3 reflectance spectra \citep[following the method described in][]{avdellidouAthorAsteroidFamily2022,Avdellidou2025MNRAS.539.3534A}, and subsequently classified according to the Bus-DeMeo taxonomic scheme \citep{demeoExtensionBusAsteroid2009} using the MIT online classification tool\footnote{\url{http://smass.mit.edu/cgi-bin/busdemeoclass-cgi}}.
A visual inspection was then performed to verify that the combined reflectance spectra matched the template spectra of the assigned taxonomic classes. In a few cases, when the MIT tool returned two equally plausible classifications, we selected the preferred class based on this visual comparison and on the asteroid's geometric visible albedo $p_\text{V}$, ensuring that the latter was consistent with the expected range for the given spectroscopic class.

\section{Results}
\label{S:results}
The results of our new observations and the spectral classification are presented in Figs.~\ref{F:ourSpectra1}-\ref{F:ourSpectra4} and Table~\ref{T:ourClassification} respectively.
We found that 25 asteroids are classified as A, four as L, two as V, two as S, three as Srw, and 51 as Sw-types. We remind that the "w" does not represent a class in the Bus-DeMeo scheme but denotes that these objects have higher reflectance slopes within their class. The latter is somewhat expected, as we focused our IRTF observations on objects having large PC1-values, hence large spectral slopes.

\begin{figure*}[h]
   \includegraphics[width=\columnwidth]{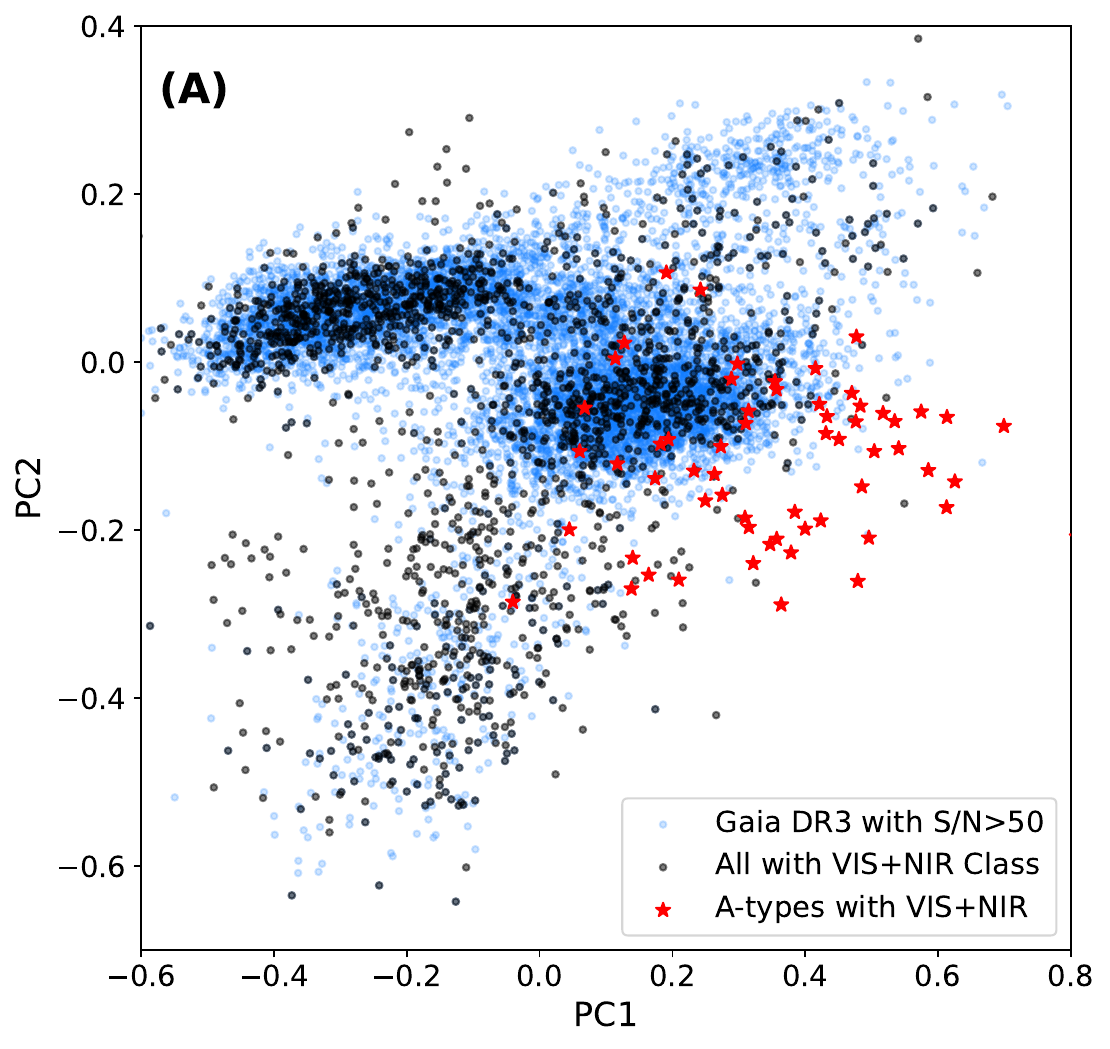}
   \includegraphics[width=\columnwidth]{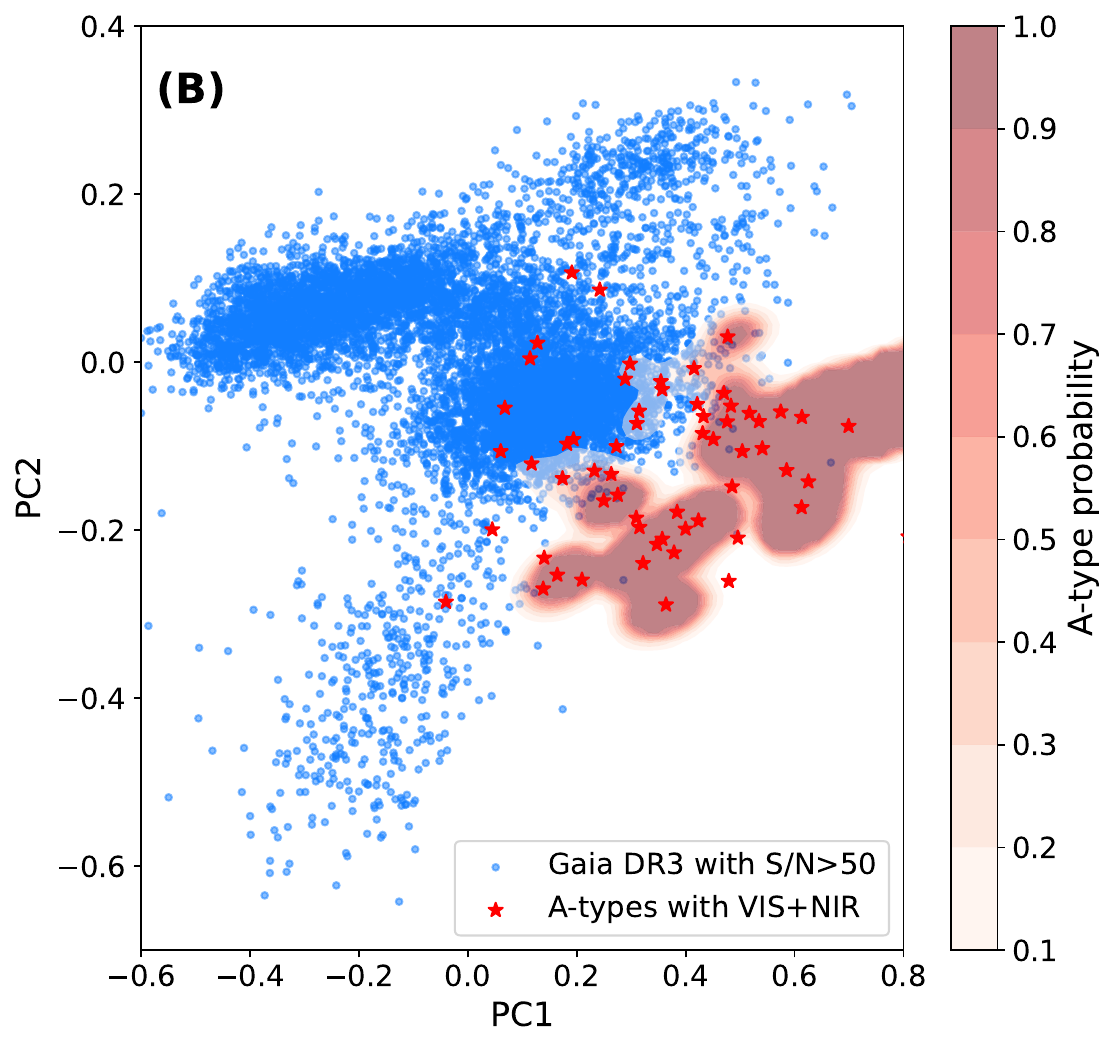}
   \caption{(A) is as Fig.~\ref{F:PCA50} but now showing A-type and non A-type asteroids with VIS+NIR spectroscopy after our survey with the IRTF. (B) shows a contour plot of the value of $K_a(x,y) / K_c(x,y)$, which approximates the  probability $P$ to find an A-type at $(x,y)$ given there is an asteroid at $(x,y)$, i.e., $P(\text{A-type}~|~(x,y))$. The KDEs were calculated here for display with $h$=0.2.}
   \label{F:AnotApca}
\end{figure*}

We combined the classification obtained in this work (Table~\ref{T:ourClassification}) for the 88 asteroids that we observed (Table~\ref{T:ourIRTFObs}), with the classification of other main belt asteroids from the literature (Table~\ref{T:literatureAtypes}) that also have Gaia DR3 reflectance spectra. 
From this combined sample we defined the set $\mathcal{S}_A$, totaling $N$~=~65 objects classified A-types. Over our IRTF runs we also observed four asteroids already known to be A-types, namely, (289) Nenetta, (1709) Ukraina, (4402) Tsunemori, and (16520) 1990 WO$_3$. In particular, 1709 was also observed by \cite{demeoOlivinedominatedAtypeAsteroids2019}. These authors obtained a spectrum similar to the A-type one after visual inspection, however they did not provide a classification for this asteroid.
The set $\mathcal{S}_{\tilde A}$ was defined by all main belt asteroids from the literature that were not classified A-type from VIS+NIR spectroscopy and the asteroids from our programme that are not classified A-type as first preferred class  (Table~\ref{T:literatureAtypes}), totalling 1,774 asteroids. The set $\mathcal{C}$ is the union of $\mathcal{S}_A$ and $\mathcal{S}_{\tilde A}$ and thus contains 1,839 asteroids. 
The positions of sets $\mathcal{S}_A$ and $\mathcal{C}$ in the PC1--PC2 plane is shown in Fig.~\ref{F:AnotApca}a, while Fig.~\ref{F:AnotApca}b shows the probability $P(\text{A-type}~|~(x,y))$ of finding an A-type at $(x,y)$ based on its PC1 and PC2 coordinates, given there is an asteroid at $(x,y)$, calculated for a particular value of the KDE bandwidth, $h=$~0.2 (see details in Appendix~A).
We applied a Monte Carlo method (see details in Appendix~A) performing 10$^4$ simulations and found a mean A-type abundance of 2.00\% with a standard deviation of 0.15\% in the main belt.

\section{Discussion}
Fig.~\ref{F:AnotApca}b shows that the highest probability of finding A-types corresponds to large PC1 and negative PC2 values, as expected from the positions in the PC1--PC2 plane of most A-types confirmed by VIS+NIR spectroscopy. However, there are some outliers: namely, two, four, and two asteroids with PC1--PC2 values in the regions of V-, S-, and L-types, respectively, with $P(\text{A-type}) \lesssim $ 20\% (compare Fig.~\ref{F:AnotApca}b and Fig. \ref{F:pcaClasses}).

\begin{figure}[h]
   \includegraphics[width=\columnwidth]{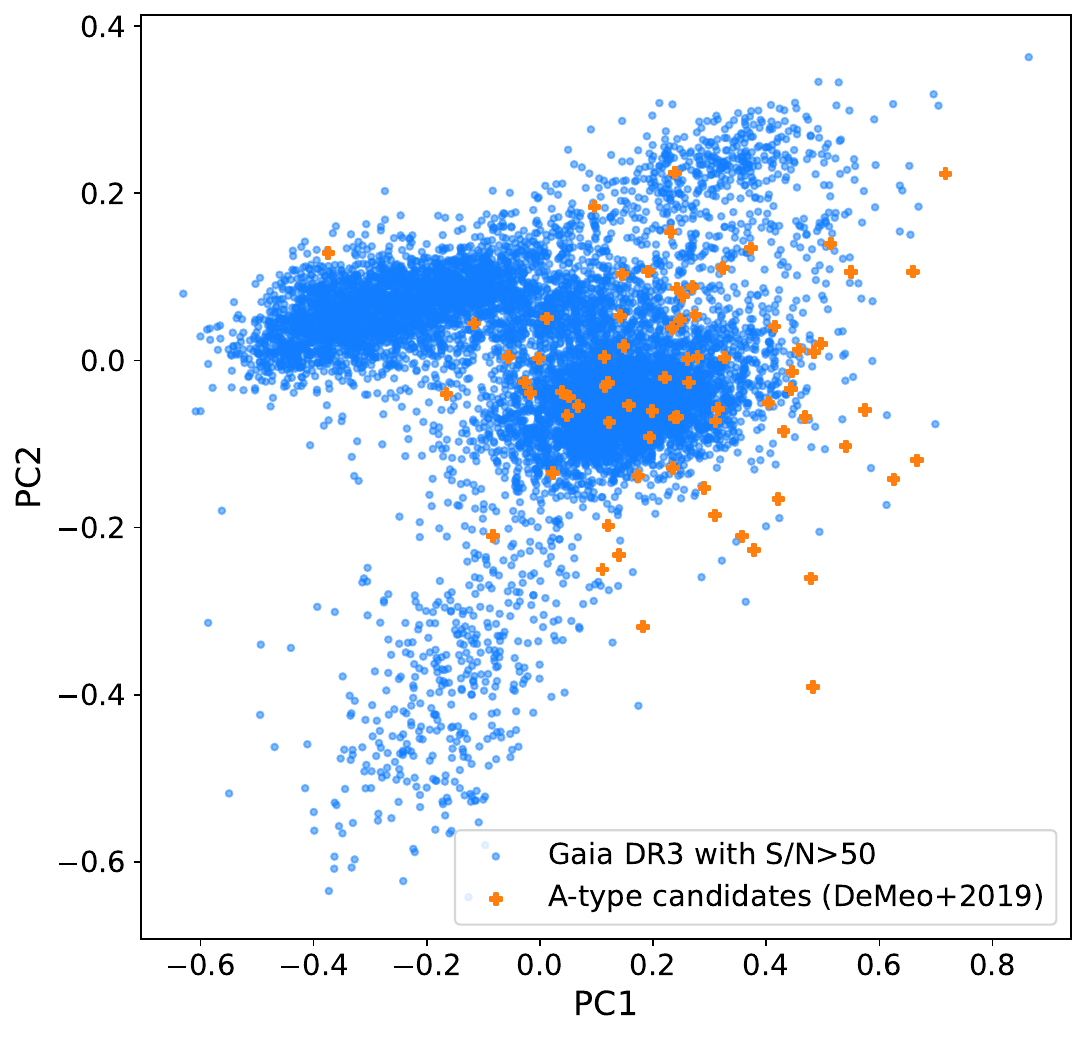}
   \caption{As in Fig.~\ref{F:PCA50} but showing the PC1, PC2 positions of the A-type candidates selected by \cite{demeoOlivinedominatedAtypeAsteroids2019} from the SDSS that have values within the Gaia DR3.}
   \label{F:DeMeoPCA}
\end{figure}

The A-type abundance we found is significantly higher than the one obtained by a previous survey \citep{demeoOlivinedominatedAtypeAsteroids2019}, but is more consistent with findings inferred from the study of NEAs \citep{popescuTaxonomicClassificationAsteroids2018, devogeleVisibleSpectroscopyMission2019,Perna2018P&SS..157...82P} and the asteroids in the Hungaria region \citep{lucasHungariaAsteroidRegion2019}. In Fig.~\ref{F:DeMeoPCA} we plot the PC1, PC2 positions, calculated from the Gaia DR3 reflectance spectra, of the A-type candidates selected by \cite{demeoOlivinedominatedAtypeAsteroids2019} using as their input the SDSS data set. When we compared these with Fig.~\ref{F:AnotApca}b, it is clear that most of the A-type candidates of \cite{demeoOlivinedominatedAtypeAsteroids2019} plot in regions of low values of the probability to find A-types, likely explaining their inferred low A-type abundance in the main belt. 

\subsection{Albedo distribution \label{S:albedo}}
We obtained the geometric visible albedos of the A-types in our list from the best-values table of the Minor Planet Physical Properties Catalogue 
\footnote{\url{https://mp3c.oca.eu} (version 3.2.1-beta.1 of 2024-08-12)}, 
which are reported in Table~\ref{T:literatureAtypes} and Table~\ref{T:ourClassification}. It is reasonable to assume that the albedo distribution is log-normal, allowing us to calculate an average geometric visible albedo of 0.26$^{+0.11}_{-0.08}$, where the asymmetric errors represent the dispersion of the distribution calculated from the symmetric 1 standard deviation of the $\log p_V$ distribution. 

\subsection{Size distribution \label{S:SFD}}
\begin{figure}[h]
   \includegraphics[width=\columnwidth]{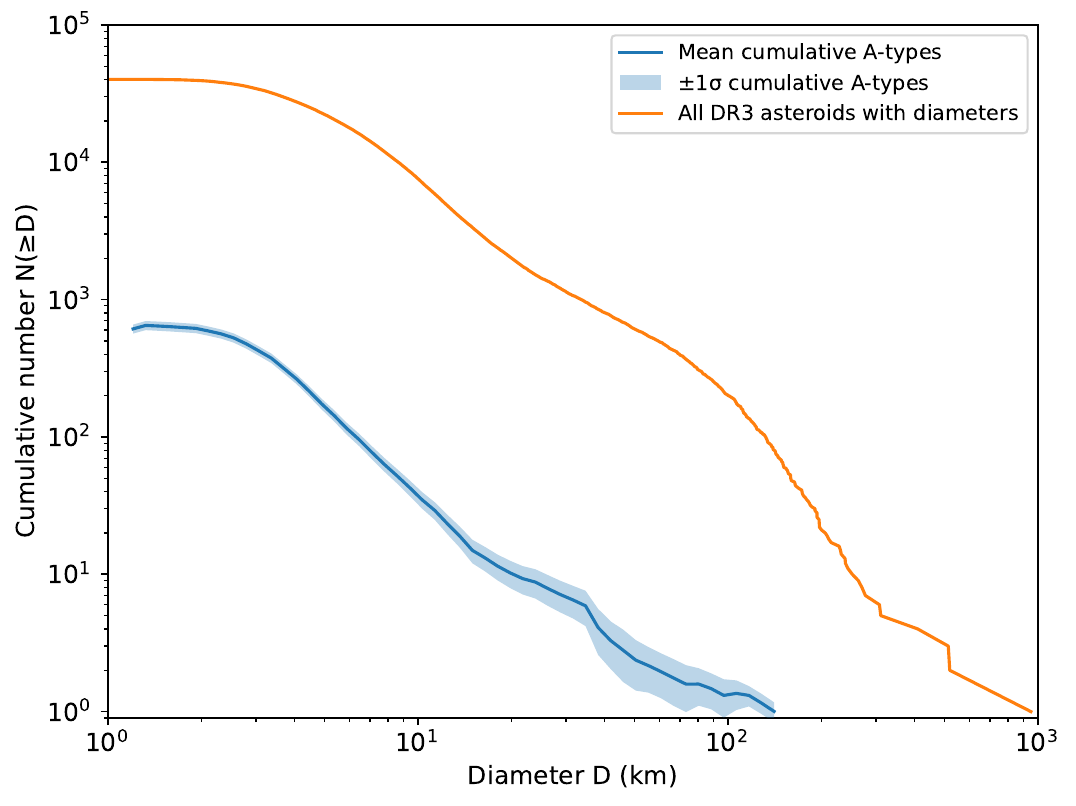}
   \caption{Size frequency distribution of A-type asteroids from the Gaia DR3 that have a known diameter, uncorrected (blue) and corrected (orange) for observational biases (see text).}
   \label{F:SFD}
\end{figure}

To build the size-frequency distribution (SFD) of main-belt A-type asteroids, we must consider that our A-type identification from the Gaia DR3 PC1--PC2 values has meaning only in a statistical sense. Therefore, we adopted a Monte Carlo approach as before: at each iteration, after evaluating $P(x_i, y_i)$ for each $(x_i, y_i)$ corresponding to a Gaia DR3 main-belt asteroid, we drew a random number uniformly distributed between 0 and 1. We identified the object as an A-type if the asteroid $P$-value was greater or equal than the random number. Then, considering only those statistically identified A-types having a known diameter, we interpolated their cumulative size-frequency distribution (SFD) at each iteration using the \emph{interp1d} function from the \texttt{scipy.interpolate} package in Python. Finally, we calculated the average and standard deviation of these series of interpolated SFDs on a grid of logarithmically spaced diameter values between 1 and 1,000~km (Fig.~\ref{F:SFD}).
We note that the average of several SFDs is not necessarily a monotonically increasing function with decreasing diameter.

However, the function calculated as detailed above does not represent the true SFD of main belt A-types, because it must be corrected for the observational biases. To do so, we calculated the ratio between the SFD \#6 of \cite{Bottke2020AJ....160...14B}, which is one of the most recent updates of the true main belt asteroids SFD, and the SFD the of Gaia DR3 main belt asteroids that have a known diameter. This ratio naturally represents the multiplicative correction that one needs to apply to the Gaia DR3 asteroids with known diameter to obtain the true SFD of the main belt. Hence, we multiplied the Gaia DR3 A-type SFD by the aforementioned ratio for each diameter bin to derive our best estimate of the true SFD of main belt A-type.We present it in Fig.~\ref{F:SFD}, where we can infer between 7,000 and 8,000 A-types with diameter $>$2~km. This is a factor $\sim$10 larger than the estimate of \cite{demeoOlivinedominatedAtypeAsteroids2019}, and it is similar to our mean A-type abundance in the main belt. 

\begin{figure*}[ht!]
   \includegraphics[width=\textwidth]{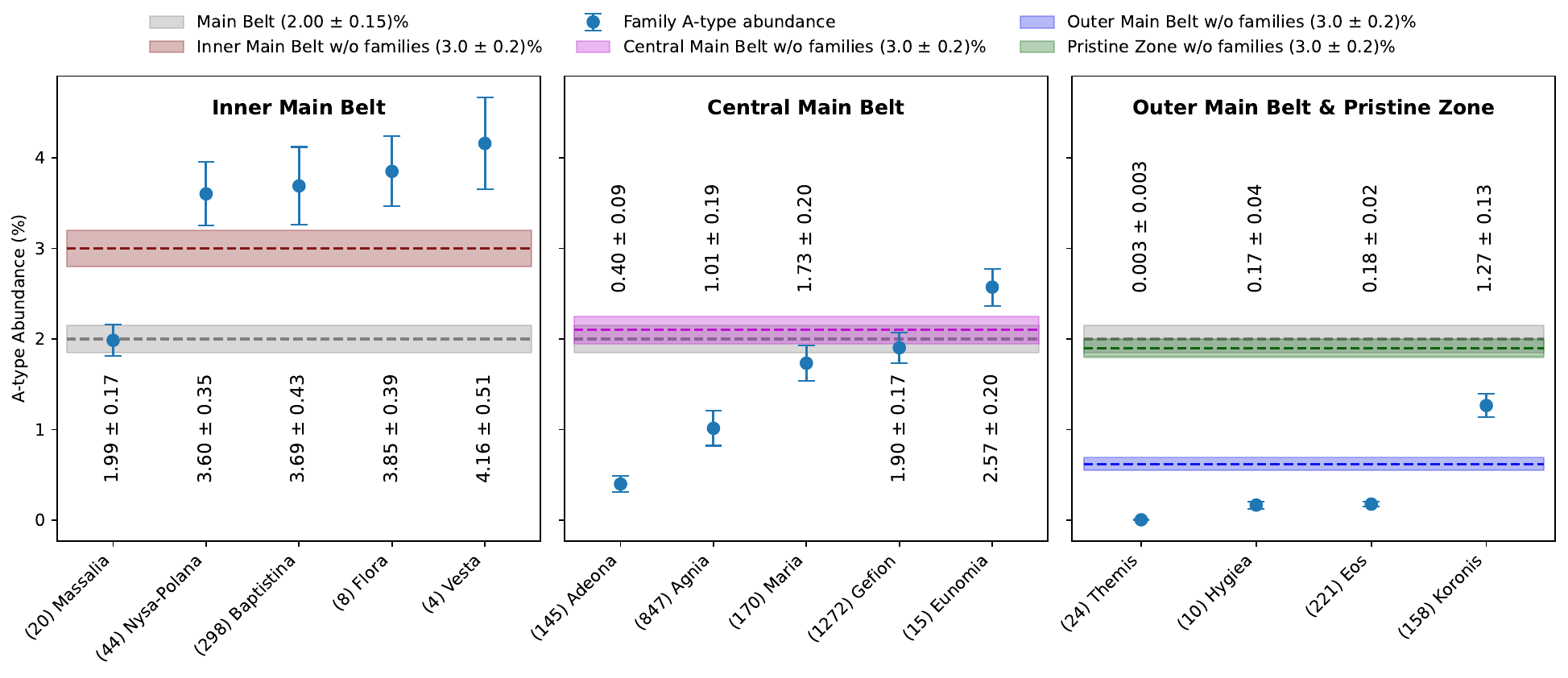}
   \caption{A-type abundance (with values) in collisional families with more than 2,000 members according to the definition of  \cite{nesvornyIdentificationFamiliesAstIV2015}, compared to the average value of the main belt and the host region of the family in the belt (inner, central, outer). Values per family, given within the plot, are in \%. The values after the $\pm$ symbol represent the standard deviation of the distribution of the A-type abundance.}
   \label{F:inFamilies}
\end{figure*}

\subsection{Heliocentric distribution}
We applied the Monte Carlo method, as described in section \ref{S:workingHypothesis}, performing 10$^4$ simulations but selecting asteroids with orbital proper semi major axis $a_p$ within specific intervals corresponding to the major sub-regions of the main belt. Namely, we considered the Inner Main Belt for $2.1 \leq a_p < 2.5$~au, the Central Main Belt for $2.5 \leq a_p < 2.82$~au, the pristine zone for $2.82 \leq a_p < 2.96$~au and the Outer Main Belt for $2.96 \leq a_p < 3.25$~au. We took proper orbital elements from the asteroid family portal\footnote{\url{http://asteroids.matf.bg.ac.rs/fam/properelements.php}} (January 2025 release), where they are computed numerically by means of a synthetic theory by \citet{Knezevic2000CeMDA..78...17K,knezevicProperElementCatalogs2003}. We found mean A-type abundances of (3.25 $\pm$ 0.25)\%, (1.91 $\pm$ 0.13)\%, (1.51 $\pm$ 0.13)\%, (0.43 $\pm$ 0.04)\% for these regions, respectively (the number after the $\pm$ is the standard deviation of the distribution). This indicates that the abundance of A-types in the main belt decreases with increasing heliocentric distance. 

When we removed the asteroid members of known collisional families, using the family membership definition from \cite{nesvornyIdentificationFamiliesAstIV2015}, we found that the mean A-type abundances in the aforementioned regions of the main belt are, respectively, (3.0$\pm$0.2)\%, (2.10$\pm$0.15)\%, (1.8$\pm$0.1)\%, and (0.62$\pm$0.07)\%, still indicating that the abundance of A-types in the main belt decreases with increasing heliocentric distance. 

\subsection{A-type abundance in collisional families}

We also applied the Monte Carlo method of section~\ref{S:workingHypothesis} with 10$^4$ iterations selecting asteroids belonging to major collisional families with more than 2,000 members. Moreover, as a control for our method, we also included the family (36256) 1999 XT17, which was discovered by \citet{galinierDiscoveryFirstOlivinedominated2024} to have a high abundance of A-type asteroids. We took family member definition from \cite{nesvornyIdentificationFamiliesAstIV2015}. The resulting A-type abundance is shown in Fig.~\ref{F:inFamilies}, while we found an A-type abundance of (14.1 $\pm$ 3.5)\% for the (36256) 1999 XT17 family (not shown in plot). The latter is consistent with previous results \citep{galinierDiscoveryFirstOlivinedominated2024} indicating that our method is robust. 

The collisional families investigated here exhibit a significant diversity in A-type asteroid abundance, with values ranging from well below the main belt average to nearly twice that level. Notably, the family of (36256) 1999 XT17 stands out with an A-type abundance more than seven times higher than the main belt average. Among the larger families, those of Vesta and Flora show the highest A-type abundances. This is not surprising. 

The Vesta family is, amongst the largest collisional families, the one with the highest A-type asteroid abundance, with a value of (4.2 $\pm$ 0.5)\%. This family consists of fragments ejected by impacts on the asteroid (4) Vesta \citep{Marchi2012Sci...336..690M}. It is well established that Vesta is a differentiated body \citep{Russel2012Sci...336..684R}, meaning that its interior separated into a core, mantle, and crust. \cite{ammannito2013} using DAWN data suggested that the presence of olivine spots detected on Vesta craters are due to impact excavation processes. However, overall there is no significant olivine on Vesta's surface. The formation of the two basins of Rheasilvia and Veneneia requires highly energetic impacts which could have excavated the crust and released olivine-rich fragments that would now be present within the family. The majority of Vesta family members exhibit spectral features similar to those of Vesta’s crustal surface, but a subset may carry signatures indicative of deeper, mantle-derived material. Subsequently, a possibility is that the basins may have been re-covered by basaltic crustal material during their modification stage or by later impacts, hiding traces of the olivine mantle.

Our analysis revealed that the Flora family ranks second among all investigated collisional families in terms of A-type asteroid abundance, with a value of (3.7 $\pm$ 0.4)\%, that is nearly double the average for the main belt. This finding strongly supports the hypothesis proposed by \citet{oszkiewiczDifferentiationSignaturesFlora2015}, who identified a significant population of V- and A-type candidates in the Flora region, potentially indicating the remnants of a differentiated parent body distinct from (4) Vesta. Their work, which used taxonomic classification based on SDSS colours, albedos, and phase-curve parameters, suggested that many of these objects cannot be dynamically linked to the Vesta family over the past 100~Myr. Our confirmation of an unusually high fraction of olivine-rich (A-type) material within the Flora family is consistent with this scenario and lends further weight to the idea that Flora may represent the fragments of another differentiated body although, contrary to Vesta, only partially.

As a caveat of the above discussion, one would expect these families to exhibit an A-type number density comparable to that of the background asteroids in the inner main belt (the region within which these families are embedded), which is estimated to be (3.0 $\pm$ 0.2)\%. Since this is not the case, we can infer that the higher-than-average A-type abundance within the Vesta and Flora families is unlikely to be due to interlopers (see also Fig.~\ref{F:inFamilies}).

\begin{figure}[h]
   \includegraphics[width=\columnwidth]{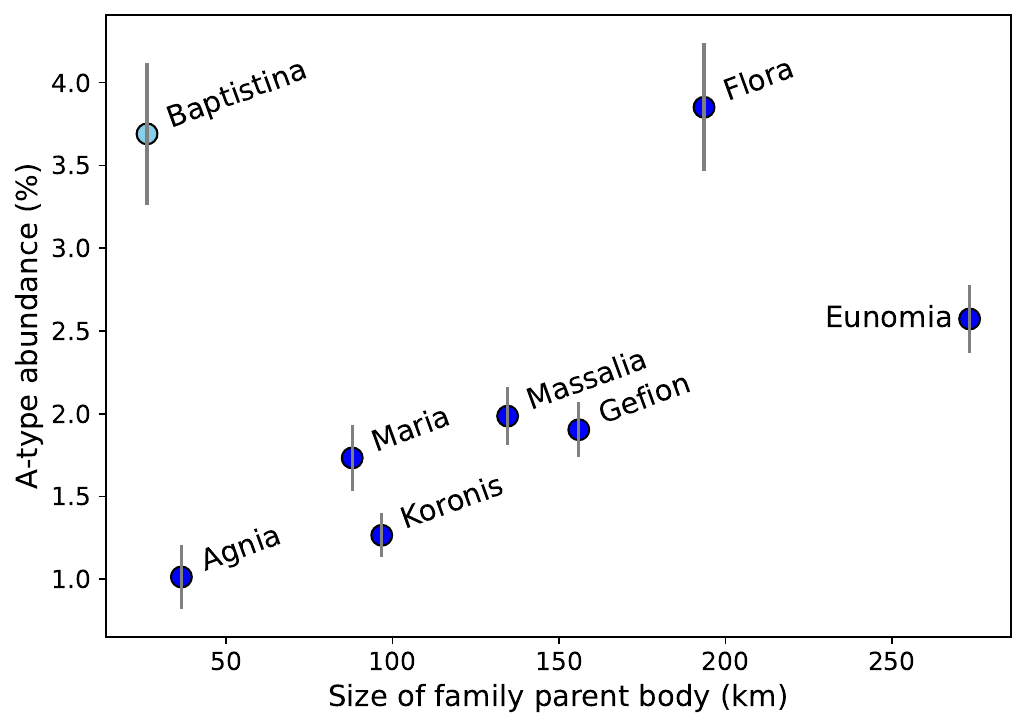}
   \caption{A-type abundance in S-complex collisional families with more than 2,000 members according to the definition of  \cite{nesvornyIdentificationFamiliesAstIV2015} as a function of the estimated diameter of the family parent body. }
   \label{F:S-typeFamilySize}
\end{figure}

Detecting differentiation in S-complex families originating from parent bodies that are, in general, considered non-differentiated -- because spectroscopically linked to the ordinary chondrites \citep[][and references therein]{brozYoungAsteroidFamilies2024,marsset2024} -- requires some considerations. If the parent body is partially differentiated this means that the crust could maintain its chondritic composition. Therefore, it should not be surprising to see differentiated and undifferentiated (chondritic) material together. Figure~\ref{F:S-typeFamilySize} shows the A-type abundance in S-complex collisional families with more than 2,000 members as a function of the estimated diameter $D_p$ of the family parent body.
The latter is calculated as $D_p = (\sum_i D_i^3 )^{1/3}$, where $D_i$ is the diameter of the $i-$th family member (only members with known diameters are included). The values of the $D_i$ are taken from the Minor Planet Physical Properties Catalogue, best values, database version: 3.3.3-beta.1 (2025-10-03). Baptistina family's symbol is light-toned because its composition is not well constrained: it is indicted as X-complex by \citet{nesvornyIdentificationFamiliesAstIV2015} and as S-complex by \citet{reddyComposition298Baptistina2009}. In addition, the Baptistina family is strongly embedded into the Flora family: indeed, the two families show similar A-type abundance. Neglecting Baptistina, Fig.~\ref{F:S-typeFamilySize} shows a correlation (Pearson correlation $\sim$0.7) between the A-type abundance in S-complex families and the size of their parent bodies, with larger parent bodies exhibiting higher A-type abundances. This could be an indication that larger S-complex parent bodies were more likely to have undergone partial differentiation, than smaller ones, which is logical as larger bodies retain more heat than smaller ones \citep[e.g.,][]{trieloffEvolutionParentBody2022}. However, the process of differentiation is complex and depends on various factors, including the body's size, composition, as well as time of the accretion, with the latter strongly controlling the level of internal heating and (partial) differentiation. For example, a body could have started accretion early, within the first 1.5~Myr after the formation of calcium-aluminum-inclusions (CAIs), then undergone differentiation and then continue its accretion at later times forming chondritic undifferentiated layers \citep{weissDifferentiatedPlanetesimalsParent2013}. Moreover, planetesimals can be implanted into the main belt already in a fragmented state from other regions of the solar system \citep{avdellidouAthorAsteroidFamily2022,avdellidouDatingSolarSystems2024}. This could also be a possibility to explain the A-type family (36256) 1999 XT17 \citep{{galinierDiscoveryFirstOlivinedominated2024}}, which is very small. Subsequent family-forming collision in the main belt could have excavated varying amounts of crust and mantle material in order to produce the observed compositional distributions within the families.

\section{Conclusions}
We performed a statistical and spectroscopic analysis of A-type asteroids in the main belt by combining Gaia DR3 VIS reflectance spectra with ground-based NIR observations. Our classification pipeline identified a population of A-type candidates, and we confirmed several of them with IRTF observations. We combined our observations with literature data, allowing us to derive a new estimate of the A-type abundance across the main belt, which we found to be ($2.00 \pm 0.15$)\%, a significantly higher fraction than previous studies had suggested.

By examining the abundance of A-types within known collisional families, we discovered a marked diversity: some families, such as Vesta and Flora, exhibit above-average A-type fractions, whereas others, such as Themis and Hygiea, display an almost complete absence of this asteroid type. The Flora family, in particular, stands out as the second most A-type-rich large family, in line with the results of \citet{oszkiewiczDifferentiationSignaturesFlora2015}, who hypothesised the existence of a second differentiated parent body in the inner main belt in addition to (4) Vesta.

These findings lend new observational support to the idea that olivine-rich materials—and by extension, fragments of differentiated bodies—are more common in the asteroid belt than previously thought. This suggests that either differentiation was more widespread in the early solar system or that our past observational constraints were incomplete. Our work thus contributes to resolving the long-standing Missing Mantle Problem and opens new avenues for identifying remnants of early planetary building blocks.

This work highlights the scientific potential of combining visible-light data from Gaia DR3 with NIR spectrophotometry for the taxonomic classification of asteroids. In particular, it demonstrates the value of broad-wavelength coverage for identifying and confirming rare asteroid types, such as A-types. Looking forward, NASA's SPHEREx mission is expected to deliver NIR spectrophotometry for a vast number of solar system objects \citep{Lisse2024AGUFMP43F...02L}. When combined with Gaia DR3 reflectance spectra, SPHEREx will enable the compilation of a powerful, multi-band spectral dataset that will allow for systematic and robust taxonomic classification of hopefully most of the asteroids present in Gaia DR3. Our results anticipate the potential of such a dataset both to improve the estimation of Eq.\ref{E:proba} across different taxonomic classes and to enable the direct identification of rare classes, thereby providing a preview of the transformative science it will make possible.

\begin{acknowledgements}
We acknowledge support from the ANR ORIGINS (ANR-18-CE31-0014) and the French Space Agency CNES. UB acknowledges funding from an STFC PhD studentship. This work has made use of data from the European Space Agency (ESA) mission Gaia (\url{https://www.cosmos.esa.int/gaia}), processed by the Gaia Data Processing and Analysis Consortium (DPAC, \url{https://www.cosmos.esa.int/web/gaia/dpac/consortium}). Funding for the DPAC has been provided by national institutions, in particular the institutions participating in the Gaia Multilateral Agreement. This work is based on data provided by the Minor Planet Physical Properties Catalogue (MP3C) of the Observatoire de la C\^ote d'Azur. CA and MD were Visiting Astronomers at the Infrared Telescope Facility, which is operated by the University of Hawaii under contract 80HQTR19D0030 with the National Aeronautics and Space Administration.
\end{acknowledgements}

\bibliographystyle{aa}
\bibliography{references.bib}

\begin{appendix}
\section{A-type probability estimation}
\label{A:appendix}
Having the two sets of asteroids, namely, A-type and non A-type from VIS+NIR spectroscopy from the literature and our own survey, we defined the two sets of asteroids, i.e., the 
$\mathcal{S}_A = \{ (x_i, y_i) \}_{i=1}^N$ and the
$\mathcal{S}_{\tilde A} = \{ (x_j, y_j) \}_{j=1}^{\tilde N}$,
where $i$ and $j$ are indexes running on the set of the $N$ asteroids classified as A-type and on the $\tilde{N}$ non A-types ones.
We then defined the $\mathcal{C}$ set to be the union of the previously two defined sets, namely, 
$\mathcal{C} = \mathcal{S}_A \cup \mathcal{S}_{\tilde A}$. Since the sets  $\mathcal{S}_A$ and $\mathcal{S}_{\tilde A}$ are disjoint, i.e., $\mathcal{S}_A \cap \mathcal{S}_{\tilde A} =\emptyset$, then the total number of elements in the union is $ |\mathcal{C}| = | \mathcal{S}_A \cup \mathcal{S}_{\tilde A} | = N + \tilde{N}$. 

We also defined the distributions $a(x,y) = \sum^{N}_i \delta(x-x_i, y-y_i)$ and  
$\tilde{a}(x,y) = \sum^{\tilde{N}}_{j} \delta(x-x_j, y-y_j)$,
where $i$ and $j$ are indexes running on the set $\mathcal{S}_A $ and the set $\mathcal{S}_{\tilde A}$, respectively, while $\delta(x-x_0,y-y_0)$ is the Dirac delta function that is non-zero only at $x=x_0$,$y=y_0$ and has the property that $\iint \delta(x-x_0,y-y_0) dx dy = 1$. 
Having noted that $\iint a(x,y) dx dy = N$ and $\iint \tilde{a}(x,y) dx dy = \tilde{N}$, one could be tempted to assume that the probability $P$ to have an A-type at $(x,y)$ (given that there is an asteroid at $(x,y)$), i.e., 
\begin{equation}
P(\text{A-type}~|~(x,y)) = \frac{\text{number density of A-type asteroids at}~(x.y)} {\text{number density of classified asteroids at}~(x,y)}
\label{E:proba}
\end{equation}
can be given by $a(x,y)/c(x,y)$ where $c(x,y) = a(x,y) + \tilde{a}(x,y)$ represents all the asteroids that have an A-type class together with those that have a class that is not A-type, both classified from VIS+NIR spectroscopy. Since both $a(x,y)$ and $c(x,y)$ are sums of Dirac delta functions, this ratio is not a smooth function, but is rather defined only at locations where $c(x,y) \neq 0$. Namely, if $(x,y)$ corresponds to an A-type, then $a(x,y) = \delta$ and $c(x,y) = \delta$, hence $a(x,y)/c(x,y)=1$, whereas when $(x,y)$ corresponds to a non A-type, then $a(x,y) = 0$ and $c(x,y) = \delta$, hence, $a(x,y)/c(x,y)=0$. But if $(x,y)$ does not correspond to any objects represented by the $c(x,y)$ function the ratio $a(x,y)/c(x,y)$ is undefined. The latter situation is, precisely, our case, because we aim at evaluating $P(\text{A-type}~|~(x,y))$ at those points where Gaia DR3 asteroids exist, which is a more numerous set of objects than those represented by the $c(x,y)$ function. 

A common solution to this problem is to smooth the sum of Dirac delta distributions by a kernel function of the form $k(x-x_i,y-y_i) = \exp \left [- \left ((x-x_i)^2 + (y-y_i)^2 \right )/(2h^2)\right]$, where $h$ is the so-called smoothing factor (also known as bandwidth). To implement this method, we defined:
\begin{align}
K_a(x,y)  & =  \frac{1}{(h^2 2 \pi)} \sum_{i=1}^{N} k(x-x_i, y-y_i)~~\text{and} \\ 
K_c(x,y)  & =   K_a(x,y)  + \frac{1}{(h^2 2 \pi)} \sum_{j=1}^{\tilde N} k(x-x_j, y-y_j)
\end{align}
to represent the Kernel Density Estimations (KDEs) of the PC1--PC2 values of the set $\mathcal{S}_A $ of the A-types, and the $\mathcal{S}_C$ set of non A-types together with the A-types (from VIS+NIR spectroscopy), respectively. By construction, $\iint K_a(x,y) dx dy = N$ and $\iint K_c(x,y) dx dy = N+ \tilde N$, hence, $K_a(x,y) / K_c(x,y) \sim P(\text{A-type}~|~(x,y))$ approximates the local probability of finding an A-type asteroid at location $(x,y)$. The abundance of A-types within the Gaia DR3 can thus be estimated by calculating the quantity:
\begin{equation}
N_{\text{A}} / N_\text{MB} = 1/ N_\text{MB} \sum_{k=1}^{N_\text{MB}} K_a(x_k,y_k) / K_c(x_k,y_k), 
\label{E:AtypeAbundance}
\end{equation}
where the index $k$ runs over the set of $N_\text{MB}$ asteroids within the main belt that have PC1, PC2 values equal to $(x_k,y_k)$, respectively.

We used the \emph{KernelDensity} Python class from the \emph{scikit-learn} project \citep{pedregosaScikitlearn2011} for the estimation of the $K_a(x,y)$ and $K_c(x,y)$ functions. Instead of attempting to find the most suitable values for the smoothing factor, we used its properties to estimate the uncertainty on the value of $K_a(x,y) / K_c(x,y)$. 
Namely, we performed a Monte Carlo simulation where at each iteration a random value of the smoothing factor was extracted from a uniform distribution between 10$^{-1}$ and 10$^{0}$, with these limits corresponding to cases that where determined visually to overfit and under-fit the data, respectively. At each iteration, we calculated $K_a(x,y)$ by summing on $N - \sqrt{N}$ $(x_i, y_i)$ points randomly extracted without repetition from the set $\mathcal{S}_A$ and the function $K_c(x,y)$ by summing on $N + \tilde N - \sqrt{N + \tilde N}$ $(x_j, y_j)$ points also randomly extracted without repetition from the set $\mathcal{C}$. So, at each iteration we evaluated the ratio  $K_a(x,y) / K_c(x,y)$ to approximate $P(\text{A-type}~|~(x,y))$ and then $N_{\text{A}} / N_\text{MB}$ from Eq.\ref{E:AtypeAbundance}.

\onecolumn

\section{Literature A-type asteroids}
\begin{table*}[h]
\caption{Forty-four main belt and Mars-crosser asteroids classified as A-types in the literature using their VIS and NIR part of the spectrum that are also included in Gaia DR3.}
\label{T:literatureAtypes}
\begin{tabular}{l|l|ccccc|c}
\hline\hline
Asteroid & Family & $a$ & $e$ & sin($i$) & $p_\text{V}$ & $\sigma$ $p_\text{V}$ & Reference \\
\hline
113	Amalthea&	Flora& 2.37786& 0.08556& 0.08785& 0.235& 0.0092& 1\\
246	Asporina&	-& 2.69391& 0.10849& 0.26928& 0.166& 0.0088& 2,3\\
289	Nenetta&	-& 2.87556& 0.20191& 0.11634& 0.270& 0.0369&2\\
354	Eleonora&	-& 2.79959& 0.11149& 0.31501& 0.192& 0.0093&2,3\\
446	Aeternitas& -& 2.78836& 0.12512& 0.18423& 0.269& 0.0076&2\\
863	Benkoela& -& 3.20099& 0.02764& 0.42912& 0.367& 0.0052&2\\
984	Gretia& -&	2.80521&	0.19537&	0.15804&	0.346& 0.0191&1\\
1122	Neith	 & -&	2.60807&	0.25508&	0.08248&	0.371& 0.0204&4\\
1709	Ukraina& -& 2.37907& 0.21338& 0.13146& 0.160& 0.0198&1\\
2501 Lohja& -& 2.42118& 0.19785& 0.05753& 0.244& 0.0317&2\\
3819	Robinson& -& 2.77292& 0.14012& 0.19279& 0.358& 0.2982&5\\
4125	LewAllen& -& 1.92131& 0.11818& 0.34926& 0.160& 0.0216&6\\
4142	Dersu-Uzala& -& 1.91166& 0.15111& 0.44611& 0.224& 0.0288&7\\
4402	Tsunemori& -& 2.89017& 0.02267& 0.13426& 0.367& 0.0760&1\\
4490	Bambery&	 -& 1.93084& 0.09229& 0.44028& 0.205& 0.0055&6\\
5641	McCleese	&-& 1.81995&	0.12622&	0.37782&	0.185& 0.0208&7\\
6067	1990 QR11& -&	3.16302&	0.04871&	0.18186&	0.256& 0.0605&5\\
7468	Anfimov&	Anfimov&	3.04235&	0.12040&	0.07634&	0.257& 0.0391&5\\
7579	1990TN1&	Hungaria& 1.97822&	0.06583&	0.29201&	0.518& 0.1150&8\\
8838	1989 UW2& -&	3.14202&	0.00745&	0.19324&	0.246& 0.0195&5,6\\
9983	Rickfienberg& -& 2.70619& 0.11650& 0.14464& 0.165& 0.0328&5\\
10715	Nagler& -& 2.63329& 0.27376& 0.30364& 0.309& 0.0399&5,6\\
10977	Mathlener& -& 2.23745&	0.02405&	0.02413&	0.327& 0.0922&6\\
16520	1990 WO3&	Flora& 2.27497& 0.16065& 0.08567& 0.325& 0.1776&5,6\\
17818	1998 FE118& -& 2.77003& 0.05269& 0.15903& 0.217& 0.0400&5,6\\
17889	Liechty&	-& 2.40783& 0.10558& 0.08096&	0.460& 0.0524&5,6\\
18853	1999 RO92&	Anfimov&	3.06203&	0.09723&	0.06893&	0.302& 0.0494&5,6\\
19652	Saris&	Hoffmeister& 2.77542& 0.06044& 0.07191&	0.242& 0.0100&5,6\\
21809	1999 TG19& -& 2.69442& 0.01652& 0.06401& 0.168& 0.0356&5,6\\
23615	1996 FK12& -& 1.96056& 0.10049& 0.38565& 0.223& 0.0230&5\\
31393	1998 YG8& -& 2.19463& 0.01918& 0.06153& 0.467& 0.1164&5,6\\
33763	1999 RB84& 1999 XT17& 2.94534& 0.07571& 0.18242& -& -&9\\
34969	4108T-2&	Flora	&2.26635& 0.09135& 0.08256& -& -&6\\
35925	1999 JP104& -& 2.43812& 0.10853& 0.19865& 0.176& 0.0201&5,6\\
36256	1999 XT17& 1999 XT17& 2.93727& 0.12373& 0.19056& 0.186& 0.0330&5,6\\
52228	Protos& -& 3.20469& 0.12401& 0.46842& 0.271& 0.1763&6\\
52726	1998 GY6& &	2.84758&	0.03831&	0.30342&	-& -&5\\
60631	2000 FC26& -&	2.71579&	0.11666&	0.18269&	0.218& 0.0736&5,6\\
75810	2000 AX244& -& 2.47639& 0.10275& 0.06741& 0.275& 0.0674&5,6\\
92516	2000 ND25& -&	2.18750&	0.07058&	0.10843&	-& -&5,6\\
95560	2002 EX98& -&	2.64959&	0.11952&	0.38285&	0.294& 0.0754&5,6\\
105840	2000 SK155& -& 3.07764& 0.23761& 0.10689& -& -&5,6\\
139045	2001 EQ9& - & 2.30958&	0.13943&	0.07845&	-& -&6\\
142040	2002 QE15& -&	1.66706&	0.34452&	0.47337&	0.340& 0.0400&10,11\\
\hline
\end{tabular}
\tablefoot{The reported semimajor axis ($a$), eccentricity ($e$), sin of the inclination ($i$) as well as the uncertainty-weighted geometric visible albedo ($p_{V}$) with this uncertainty were retrieved from Minor Planet Physical Properties Catalogue. The family membership is taken from \cite{nesvornyIdentificationFamiliesAstIV2015}. The reference corresponds to the study that reported the A-type classification as follows: 
(1)~\cite{mahlkeAsteroidTaxonomyCluster2022}, (2)~\cite{demeoExtensionBusAsteroid2009}, (3)~\cite{Gietzen2012M&PS...47.1789G}, (4)~\cite{Fornasier2011Icar..214..131F}, (5)~\cite{Hasegawa2024AJ....167..224H}, (6)~\cite{demeoOlivinedominatedAtypeAsteroids2019}, (7)~\cite{Binzel2004P&SS...52..291B}, 
(8)~\cite{Fornasier2008Icar..196..119F}, (9)~\cite{galinierDiscoveryFirstOlivinedominated2024}, (10)~\cite{deLeon2010A&A...517A..23D}, (11)~\cite{Lazzarin2005MNRAS.359.1575L}.}
\end{table*}

\clearpage
\section{Observations of candidate A-types and results}
\begin{table*}[h]
\caption{Observational circumstances for NIR spectroscopy of the 88 asteroids of our programme.}
\label{T:ourIRTFObs}
\begin{tabular}{l|lll||l|lll}
\hline\hline
Asteroid & Date (UT)	&	 SA	      &	 SA & Asteroid & Date (UT)	&	 SA	      &	 SA \\
	     &			    &	trusted	&	local & 	     &			    &	trusted	&	local\\			
\hline
254	 & 2021-12-21  & SA102-1081	& HD73708 &  7752	 & 2022-07-04 & SA107-684 &	HD140990\\
281	 & 2022-02-25 &   SA98-978    &   - & 8404 & 2021-12-21 & SA115-271  &	HD202410\\
289	 & 2022-07-03 &   SA107-684  & TYC 272-939-1 & 8508	 & 2022-02-25 &  SA98-978  &	-\\
440	 & 2022-07-25	&   SA107-684   &  HD153190 & 8660 & 2022-02-25 &  SA102-1081  &	BD+19 2318\\
684	 & 2022-02-26  &   SA107-684   & HD121867 & 10017 & 2022-03-27 &   SA102-1081 &  -  \\
736	 & 2022-02-26  &   SA107-684   & HD117500 & 10212 & 2022-07-04 &   SA107-684  & HD133011\\
813	 & 2021-12-21	&   SA93-101    &  -  & 10811 & 2021-12-21 &   Hyades 64 &  HD252943  \\
1019	 & 2022-03-19 &  SA107-998   & NGC5272-1549 & 11160 & 2022-06-10  &	SA107-998 &  HD140990\\
1153	 & 2024-01-28 &	SA102-1081  &  -  & 11549 & 2022-06-10 &  SA102-1081  &   HD115106\\
1185	 & 2022-03-20 &  SA98-978  	& HD56513   & 11861 & 2021-08-21 &   SA115-271 & HD224693\\
1188	 & 2021-12-21	 &  Hyades 64	&  -  & 12366 & 2022-06-10 &  SA107-998  &  HD140990\\
1393	 & 2022-02-26 &  Hyades 64    &  -  & 12440 &	2024-08-02 & 	SA107-684 & BD124517\\
1488	 & 2021-08-24 &	SA110-361     &  HD207398 & 12701 & 2022-02-25 &  SA102-1081 &  BD+132231\\
1536	 & 2021-12-21 &	Hyades 64 &  -  & 12745 & 2021-08-24 &  SA115-271  &   HD211706 \\
1551	 & 2021-08-21 &	SA93-101	 &  -  & 13441 & 2024-10-13   & SA93101  &  - \\
1709	 & 2021-12-21 &  Hyades 64    & HD252943 & 14683 & 2022-07-25 &    SA110-361 &  HD175276\\
1814	 & 2022-03-20 &  SA98-978  & HD257880 &  14691 & 2022-07-05 &   SA107-684   & HD110747\\
1908	 & 2021-08-21 &  SA93-101  & HD4060 &  14901 & 2022-03-20 &  SA107-684   & HD12859\\
2074	 & 2022-07-05 &	SA107-684 &  BD+333318B & 15117 & 2021-08-22 &   SA110-361  &  HD177911\\
2088	 & 2022-02-26 &	SA102-1081	&  HD102046 &  15166 & 2022-06-10 &  SA107-998  &  -  \\
2171	 & 2021-12-21 &   Hyades 64 	&  -   & \ 16439 & 2021-08-22 &  SA93-101 & HD224251\\
2647 & 2022-02-25 &  SA102-1081  	&  HD80653 &  16520 & 2024-10-01 & SA115-271  &  -   \\
2802 & 2022-06-10 &   SA102-1081 	&  -   & 17015 & 2022-02-26 &  SA102-1081  &  HD82840  \\
3104 & 2021-08-24 &   SA93-101 & HD26257 & 19687 & 2021-08-22 &  SA115-271 & HD189499\\
3272 & 2022-03-20 &   SA102-1081 & BD+192318 & 20325 & 2021-08-24 & SA93-101  & -\\
3653 & 2022-02-25 &   SA102-1081	& HD82410  & 20406 & 2022-03-27 &  SA102-1081  &  -\\
3873 & 2022-07-03 &   SA107-684   & HD144821  & 21086 & 2022-03-24 &  SA102-1081  &  HD77730\\
4062 & 2021-12-21 &   SA115-271   &  -  & 22849 & 2022-02-26  &  SA102-1081 & -\\
4402 & 2022-03-19 &	SA107-998 &  HD130323  & 24473 & 2022-02-25  &  SA98-978   & HD76752\\
4402 & 2022-07-05 &	SA107-694 &  HD123353  & 25808 & 2024-09-11 & SA93101&	-\\
4509 & 2022-06-10 &	SA110-361 &  HD161555 & 26385 & 2021-08-22  &  SA115-271 &  HD353227\\
4720 & 2022-07-03 &	SA107-684 & HD147194 & 26879 & 2022-06-10  &  SA107-998 &  -  \\
5086 & 2022-02-25 &	SA102-1081 & HD94070 & 26960 & 2022-02-26  &  SA102-1081 &  BD+151753\\
5596	 & 2022-02-26 &	SA107-684 & HD115642 & 27929 & 2022-03-24  &  SA102-1081 &  -  \\
5719 & 2022-03-19 &	SA107-998  & HD115642 & 28858 & 2023-09-08  &  BD034598  & -    \\
5872	 & 2022-07-03 &	SA107-684  &  HD147194 & 31060 & 2023-11-03  &  SA93101   & -   \\
6344 & 2022-03-20 &	SA102-1081 & HD107766 & 32046 & 2022-07-05  &  SA107-684     &  HD140990\\
6384	 & 2022-03-27 &	SA102-1081 & HD120566 &  32458 & 2022-03-26 &  SA102-1081 & BD+152449\\
6418	 & 2022-03-26 &	SA102-1081 & HD115575 &  32772 & 2024-01-06  &	SA98-978 &  -  \\
6455 & 2022-03-19  &	SA107-998 &  TYC2532-862-1 & 44711 & 2022-03-20 &  SA110-361 & HD138779\\
6851 & 2022-02-26 &	SA102-1081 & HD101690 & 46773 & 2022-07-04 &	SA107-684 &  HD159054\\
6873 & 2022-06-10 &	SA110-361 &  HD161555 &  61180 & 2022-07-25 &  SA107-684  & HD183329\\
7086 & 2022-03-20 & SA107-684   &	HD133011 &  98335 & 2022-03-20 &  SA102-1081 &  HD99268\\
7294 & 2022-03-24 & SA102-1081  &	HD116942 &  141111 & 2021-08-24 &  SA115-271 &  HD209847\\
7709 & 2021-08-24 & SA93-101  &	HD2914  &  &  &  & \\
\hline
\end{tabular}
\end{table*}

\clearpage
\begin{longtable}{l|l|cccccc}
\caption{Physical properties and our spectral classification of the 88 asteroids observed in this work. The two classes in the last column are given by order of preference.}\label{T:ourClassification}\\
\hline\hline
Asteroid &	Family &	$a$	& $e$ &	sin($i$) & $p_\text{V}$ & $\sigma$ $p_\text{V}$ & Class. \\
\hline
\endfirsthead
\caption{continued.}\\
\hline
Asteroid &	Family &	$a$	& $e$ &	sin($i$) & $p_\text{V}$ & $\sigma$ $p_\text{V}$ & Class.\\
\hline
\endhead
\hline
\endfoot
254	Augusta &	Flora & 2.19456  & 0.12128 & 0.07867 & 0.163 & 0.0061 & Sw\\
281	Lucretia &	Flora & 2.18772 & 0.13194 & 0.09241 & 0.204 & 0.0113 & Sw\\
289	Nenetta & - & 2.87556 & 0.20191 & 0.11634 & 0.270 & 0.0369  & A\\
440	Theodora & - & 2.21094 & 0.10766 & 0.02786 & 0.279 & 0.0717 & Sw\\
684	Hildburg & - & 2.43188 & 0.03444 & 0.09623 & 0.176 & 0.0140 & Sw\\
736	Harvard & - & 2.20237 & 0.16546 & 0.07631 & 0.145 & 0.0082 & Sw\\
813	Baumeia	 &	- & 2.22340 & 0.02528 & 0.10966 & 0.247 & 0.0145 & Sw\\
1019	Strackea	 &	- & 1.91186 & 0.07108 & 0.45361 & 0.262 & 0.0206 &Sw\\
1153	Wallenbergia & - & 2.19535 & 0.16082 & 0.05818 & 0.414 & 0.0763 & Sw\\
1185	Nikko	 &	Flora & 2.23735 & 0.10533 & 0.09928 & 0.324 & 0.0356  & Sw\\
1188	Gothlandia &Flora & 2.19078 & 0.18023 & 0.08397 & 0.235 & 0.0107 & Sw\\
1393	Sofala & - & 2.43380 & 0.10963 & 0.10184 & 0.223 & 0.0460 & Sw\\
1488	Aura	 &	- & 3.03991 & 0.11348 & 0.18321 & 0.125 & 0.0075 & A\\
1536	Pielinen & - & 2.20451 & 0.19507 & 0.02679 & 0.246 & 0.0100 & Sw\\
1551	Argelander & - & 2.39489 & 0.06594 & 0.06558 & 0.232 & 0.0179 & Sw\\
1709	Ukraina & - & 2.37907 & 0.21338 & 0.13146 & 0.160 & 0.0198 & A\\
1814	Bach	 & - & 2.22584 & 0.13045 & 0.07578 & 0.200 & 0.0396 & Sw\\
1908	Pobeda & - & 2.88999 & 0.03840 & 0.08310 & 0.125 & 0.0108 & A\\
2074	Shoemaker & - & 1.79961 & 0.08186 & 0.50120 & 0.261 & 0.0307 & A, D\\
2088	Sahlia &	Flora & 2.20662 & 0.07970 & 0.09655 & 0.310 & 0.0361 & Sw\\
2171	Kiev	 &	Flora & 2.25576 & 0.16619 & 0.13078 & 0.101 & 0.0059 & Sw\\
2647	Sova	 &	Flora & 2.24372 & 0.13694 & 0.06865 & 0.310 & 0.0242 & Sw\\
2802	Weisell & - & 3.11191 & 0.12266 & 0.16701 & 0.224 & 0.0196 & Sw\\
3104	Durer	 & - & 2.96506 & 0.08655 & 0.40966 & 0.421 & 0.0407 & A\\
3272	Tillandz	 &Flora & 2.24340 & 0.09223 & 0.06844 & 0.266 & 0.0098  & Sw\\
3653	Klimishin & - & 2.23585 & 0.09873 & 0.08640 & 0.26 & 0.0227 & Sw\\
3873	Roddy & - & 1.89252 & 0.13364 & 0.39643 & 0.254 & 0.0306 & Sw\\
4062	Schiaparelli &	Flora & 2.24290 & 0.14919 & 0.12017 & 0.196 & 0.0282  & Sw\\
4402	Tsunemori & - & 2.89017 & 0.02267 & 0.13426 & 0.367 & 0.0760 & A, Srw\\
4509	Gorbatskij & - & 2.76203 & 0.27285 & 0.18992 & 0.148 & 0.0149 & L, Xe\\
4720	Tottori	 &	Flora & 2.22558 & 0.14572 & 0.10233 & 0.305 & 0.0390 & Sw\\
5086	Demin	 & - & 2.17701 & 0.11333 & 0.05920 & 0.252 & 0.0276 & Sw\\
5596	Morbidelli &	Flora & 2.17484 & 0.08445 & 0.07384 & 0.198 & 0.0219 & Sw\\
5719	Krizik	 &	Flora & 2.21801 & 0.16962 & 0.07613 & - & - & Sw\\
5872	Sugano & - & 2.24946 & 0.13337 & 0.11515 & - & - & Sw\\
6344	1993 VM &	Flora & 2.24723 & 0.11617 & 0.06668 & - & - & A, Sw\\
6384	Kervin & - & 1.93681 & 0.07710 & 0.44286 & 0.495 & 0.0559 & Sw\\
6418	Hanamigahara &Vesta & 2.25961 & 0.10716 & 0.10711 & 0.329 & 0.0810  & Sw\\
6455	1992 HE & - & 2.24029 & 0.57258 & 0.60655 & 0.227 & 0.0510 & Srw\\
6851	Chianti	 & - & 2.19524 & 0.12517 & 0.06650 & 0.290 & 0.1200 & Sw\\
6873	Tasaka	 &	Flora & 2.17289 & 0.13561 & 0.08351 & 0.322 & 0.0990 & L, Xe\\
7086	Bopp	 & - &1.90929 & 0.08750 & 0.43240 & 0.296 & 0.0502 &Sw\\
7294	Barbaraakey	 & - & 2.65814 & 0.09190 & 0.11687 & 0.296 & 0.0724 & A\\
7709	1994 RN1	 &	Flora & 2.24544 & 0.18591 & 0.08252 & 0.281 & 0.0311 & Sw\\
7752	Otauchunokai &	Flora & 2.29965 & 0.15325 & 0.06617 & 0.314 & 0.0549  &Sw\\
8404	1995 AN &	Hungaria & 1.96820 & 0.07697 & 0.31078 & 0.344 & 0.0351  & Sw\\
8508	1991 CU1	 & - & 2.28552 & 0.13339 & 0.06757 & 0.214 & 0.0261 & A, D\\
8660	Sano	 & - & 3.25870 & 0.03997 & 0.11555 & 0.260 & 0.0230 & A\\
10017	Jaotsungi	 & - & 2.18604 & 0.14638 & 0.15796 & 0.353 & 0.0352  & Srw\\
10212	1997 RA7	 & - & 2.16892 & 0.09157 & 0.04195 & - &	- & Sw\\
10811	Lau	 &	Lau & 2.93301 & 0.21850 & 0.12535 & 0.272 & 0.0237 & A\\
11160	1998 BH7	 & - & 2.23294 & 0.08407 & 0.02398 & - & - & A, D\\
11549	1992 YY	 & - & 2.53149 & 0.08793 & 0.19839 & 0.172 & 0.0097 & Sw\\
11861	Teruhime & - & 3.15722 & 0.10578 & 0.21529 & 0.435 & 0.0760 & A \\
12366	Luisapla &	Flora & 2.28579 & 0.11285 & 0.05978 & - & - & Sw\\
12440       Koshigayaboshi & - & 3.0633  & 0.0542 & 0.1924 & - & - & Sw\\
12701	Chenier	 & - & 2.33207 & 0.17036 & 0.11356 & - & -  & A, D\\
12745	1992 UL2	 & - & 2.19257 & 0.17458 & 0.05171 & 0.374 & 0.0456  & Sw\\
13441	Janmerlin& - & 2.63057 & 0.26137 & 0.20681 & 0.378 & 0.0471 & L, Xe\\
14683	Remy	 &	Flora & 2.27731 & 0.15697 & 0.08719 & 0.286 & 0.0430  & A, D\\
14691	2000 AK119 &	Eunomia & 2.61570 & 0.15030 & 0.21161 & 0.156 & 0.0130  & L, Xe\\
14901	Hidatakayama &	Eunomia & 2.69552 & 0.17155 & 0.21348 & 0.282 & 0.1320  & Sw\\
15117	2000 DA79	 & - & 2.88427 & 0.15636 & 0.10143 & 0.277 & 0.0593  & A\\
15166	2000 GX90	 &	Eunomia & 2.66570 & 0.17299 & 0.23664 & 0.326 & 0.0160  & Sw\\
16439	Yamehoshinokawa	 &Eunomia & 2.64511 & 0.09995 & 0.21172	 & 0.282 &	0.0288  & Sw\\
16520	1990 WO3	 &Flora	 & 2.27497 & 0.16065 & 0.08567 & 0.325 & 0.1776 & A\\
17015	Shriyareddy	 &Flora & 2.21540 & 0.10993 & 0.06248 & 0.326 & 0.0234 & Sw\\
19687	1999 RP199 & - & 2.72535 & 0.13561 & 0.12366 & 0.276 & 0.0322 & Undetermined\\
20325	Julianoey & - & 2.37935 & 0.07637 & 0.10622 & 0.416 & 0.0440  & Srw\\
20406	1998 QJ13	 & - & 2.92930 & 0.15186 & 0.12712 & 0.211 & 0.1281  & A\\
21086	1992 AO1	 & - & 2.41056 & 0.20972	 & 0.07097 & 0.330 & 0.0390  & A\\
22849	1999 RZ125 & - & 2.53009 & 0.12073 & 0.25621 & - & -  & Sw\\
24473	2000 UK98 &		Flora & 2.25664 & 0.11030 & 0.09572 & 0.304 & 0.0289  & Sw \\
25808	2000 CK103& - & 3.04315 & 0.10320 & 0.20371 & 0.199 & 0.0404 & A\\
26385	1999 RN20 & - & 2.15218 & 0.26900	 & 0.45527 & 0.262 & 0.0406  & Sw \\
26879	Haines & - & 2.31047 & 0.34857 & 0.36375 & 0.213 & 0.0446  & S\\
26960	Liouville & - & 2.39516 & 0.15737 &	0.11418 &-& - & Sw\\
27929	1997 FC1 & - & 2.34426 & 0.09406 & 0.12447 & - &- & V\\
28858   2000 JK59 & - & 2.3958  & 0.12171 & 0.20207 & 0.351 & 0.1126 & A\\
31060   1996 TB6 & - & 2.7265  & 0.31213 & 0.21238 & 0.38  & 0.14 & A\\
32046	2000 JR28 & - & 2.37084 & 0.21082 & 0.16912 & 0.264 & 0.1007 & Sw\\
32458	2000 SF87 & - & 2.45384 & 0.23484 & 0.21143 & 0.253 & 0.0427 & Sw\\
32772	1986 JL & - & 1.96102 & 0.09676 &	0.43127 &	0.163 & 0.0084 & A\\
44711	Carp & - & 2.25527 & 0.23516 & 0.09276 & 0.380 & 0.1100  & A\\
46773	1998 HZ12 & - & 2.25370 & 0.27654	 & 0.10046 & 0.190 & 0.0337 & S\\
61180	2000 NQ24 &	Vesta & 2.38330 & 0.14709 & 0.12626 & - & - & V\\
98335	2000 SP295 &	Flora & 2.26859 & 0.12013 & 0.12526 & 0.390 & 0.1525 & Sw\\
141111	2001XR64 & - & 2.99725 & 0.35257 & 0.03346 & - & -  & Sw\\
\hline
\end{longtable}

\begin{figure*}[htp]
   \centering
   \includegraphics[width=1\textwidth, trim=0mm 170mm 0mm 0mm, clip]{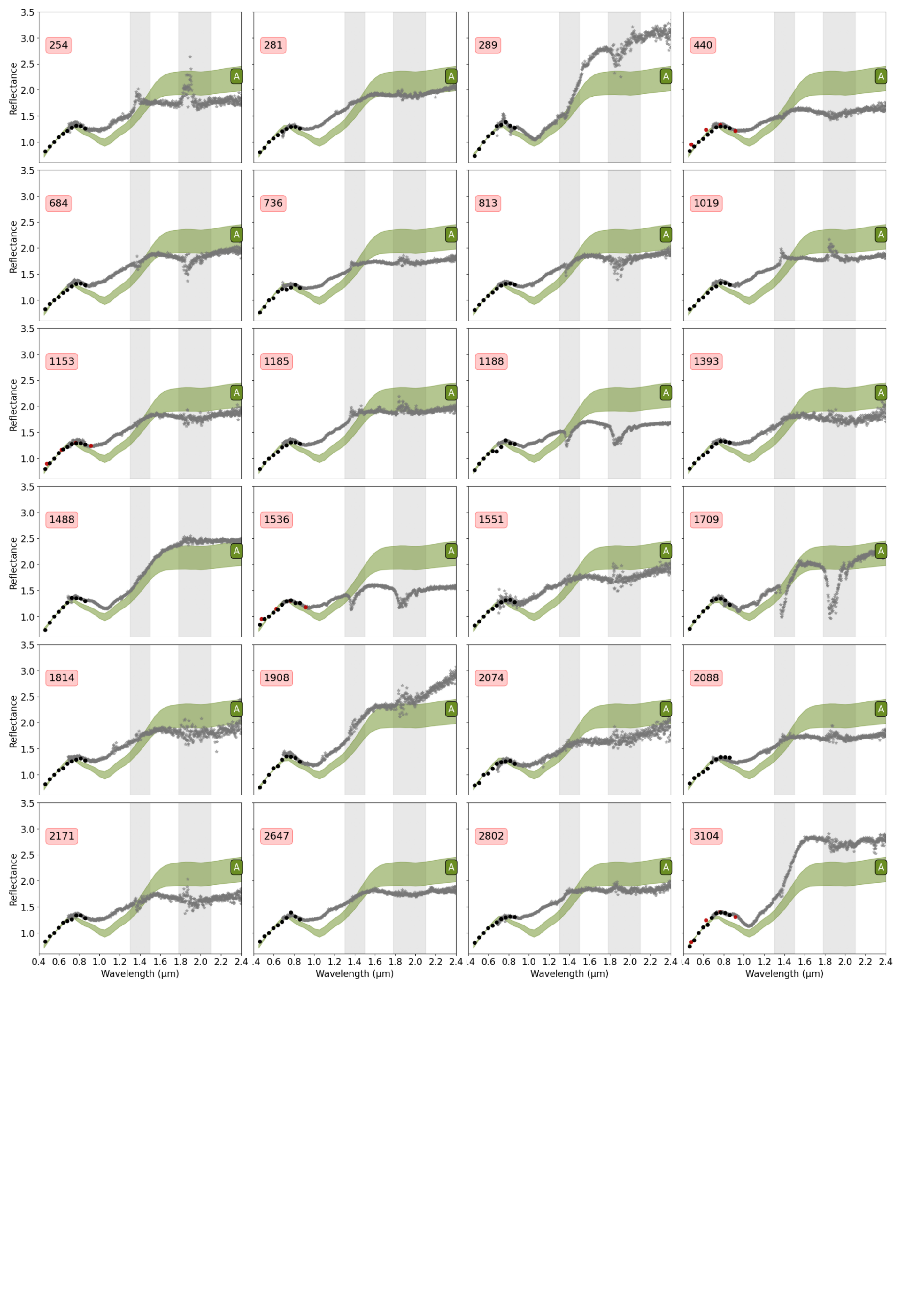}         
   \caption{Reflectance spectra of the asteroids from this work. Reflectance spectra from the SpeX instrument at the IRTF, Gaia DR3 and (when available SDSS are shown with grey stars, black filled circles and red filled circles, respectively. The green shaded areas represent the range of spectra for the A-types of the Bus-DeMeo taxonomy \citep{demeoExtensionBusAsteroid2009} for comparison to the observed reflectance spectra. The grey-shaded regions correspond to wavelength range of lower atmospheric transparency. Data in those regions, should be interpreted with care. Asteroids numbers from 254 to  3104 are shown here.} \label{F:ourSpectra1}
\end{figure*}

\begin{figure*}[htp]
   \centering
   \includegraphics[width=1\textwidth, trim=0mm 170mm 0mm 0mm, clip]{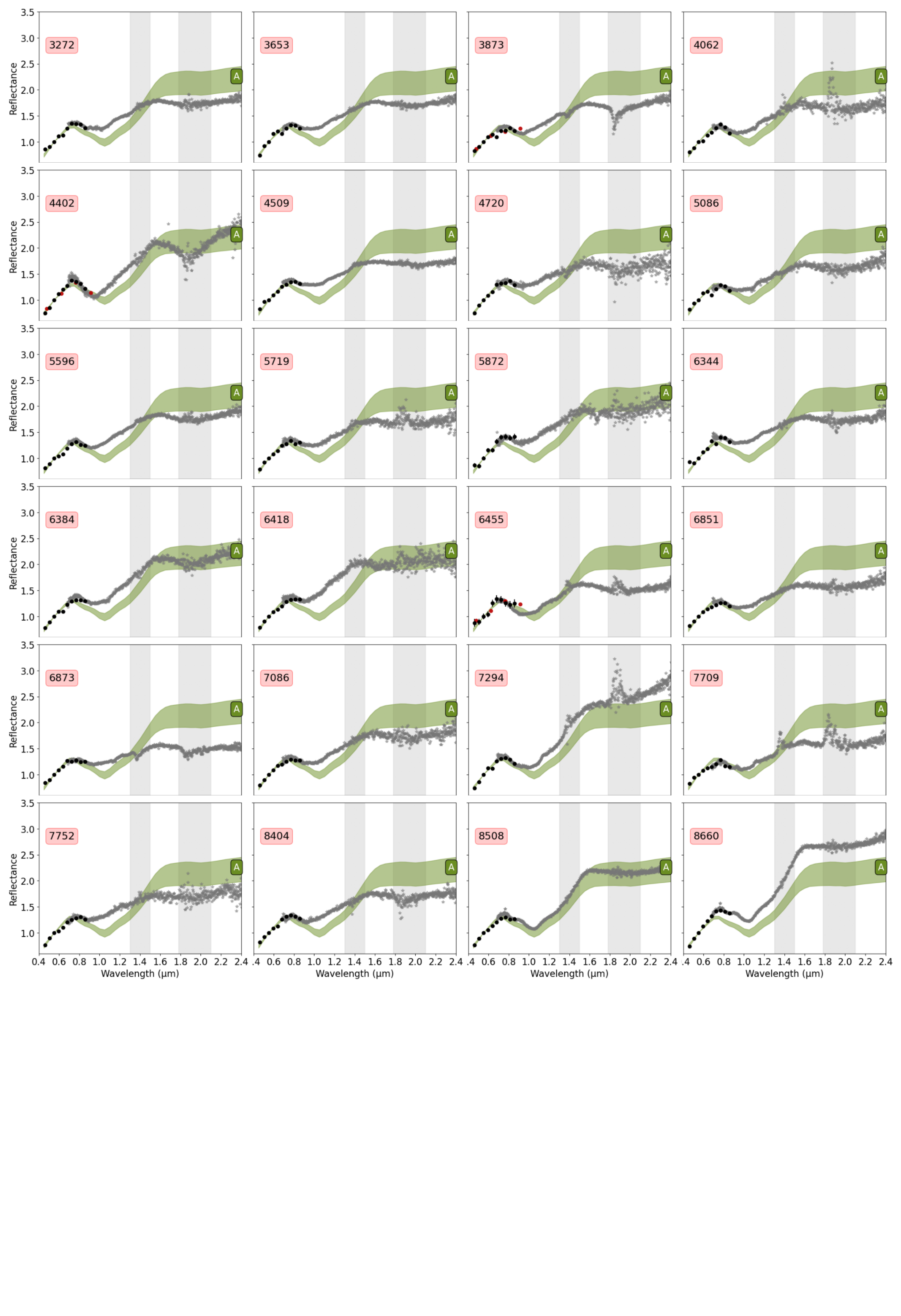}         
   \caption{Reflectance spectra of the asteroids from this work. Asteroids numbers from 3272 to  8660 are shown here. See caption of Fig.~\ref{F:ourSpectra1} for further details.}  \label{F:ourSpectra2}
\end{figure*}

\begin{figure*}[htp]
   \centering
   \includegraphics[width=1\textwidth, trim=0mm 170mm 0mm 0mm, clip]{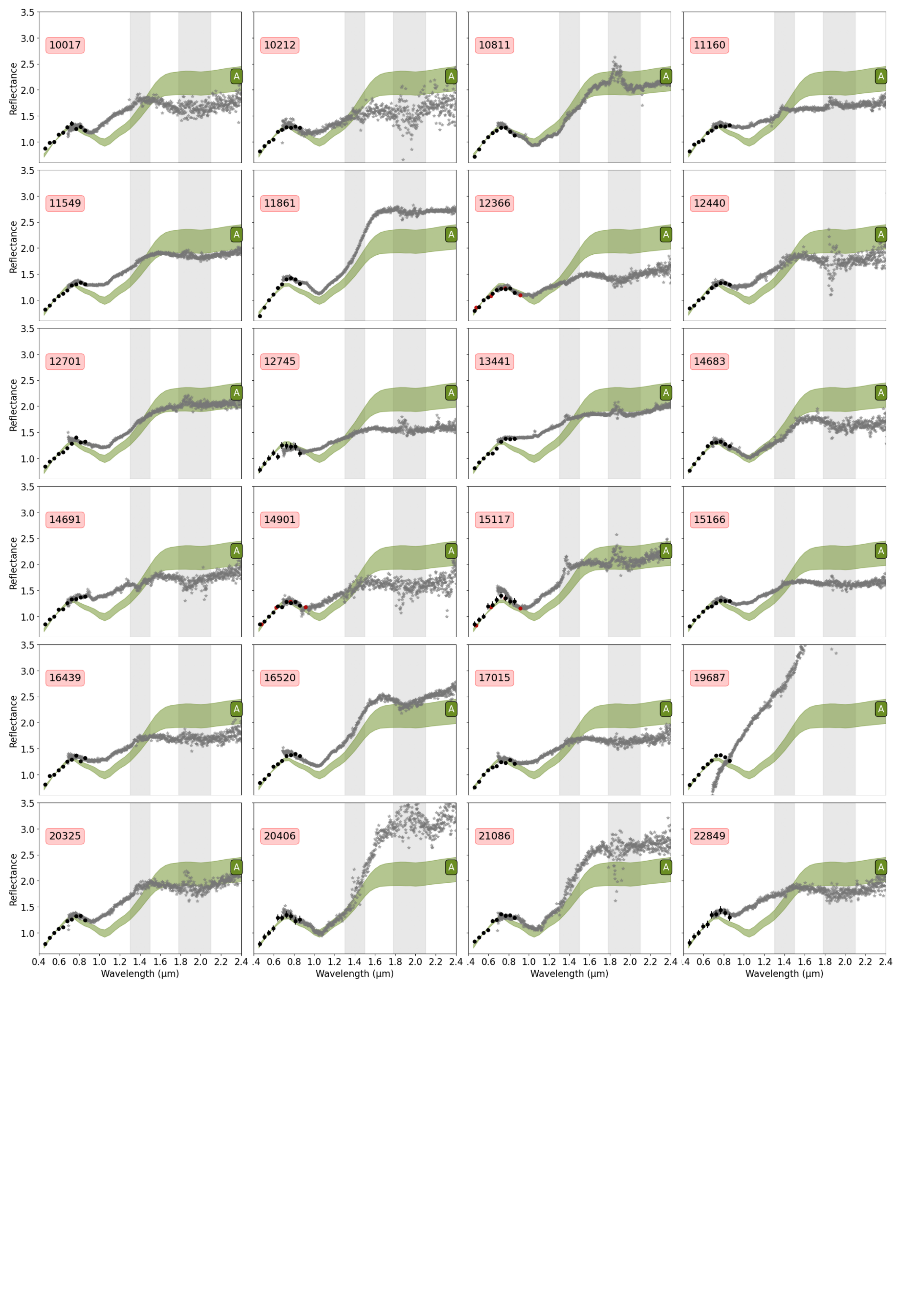}         
   \caption{Reflectance spectra of the asteroids from this work. Asteroids numbers from 10017 to 22849 are shown here. See caption of Fig.~\ref{F:ourSpectra1} for further details.}  \label{F:ourSpectra3}
\end{figure*}

\begin{figure*}[htp]
   \centering
   \includegraphics[width=1\textwidth, trim=0mm 330mm 0mm 0mm, clip]{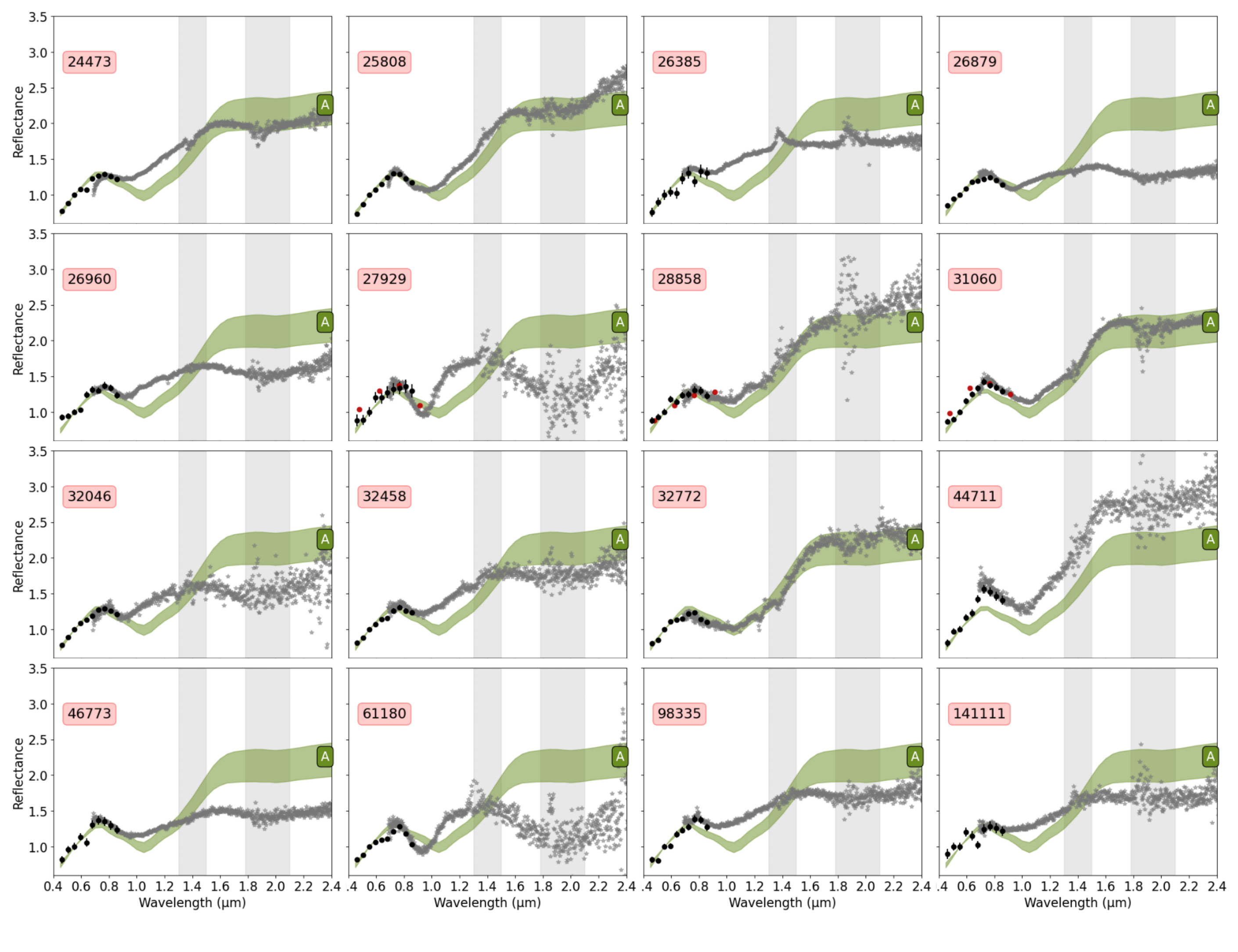}         
   \caption{Reflectance spectra of the asteroids from this work. Asteroids numbers from 24473 to 141111 are shown here. See caption of Fig.~\ref{F:ourSpectra1} for further details.}  \label{F:ourSpectra4}
\end{figure*}

\clearpage
\section{Distribution of main spectral types}
\begin{figure*}[htp]
\includegraphics[width=\textwidth]{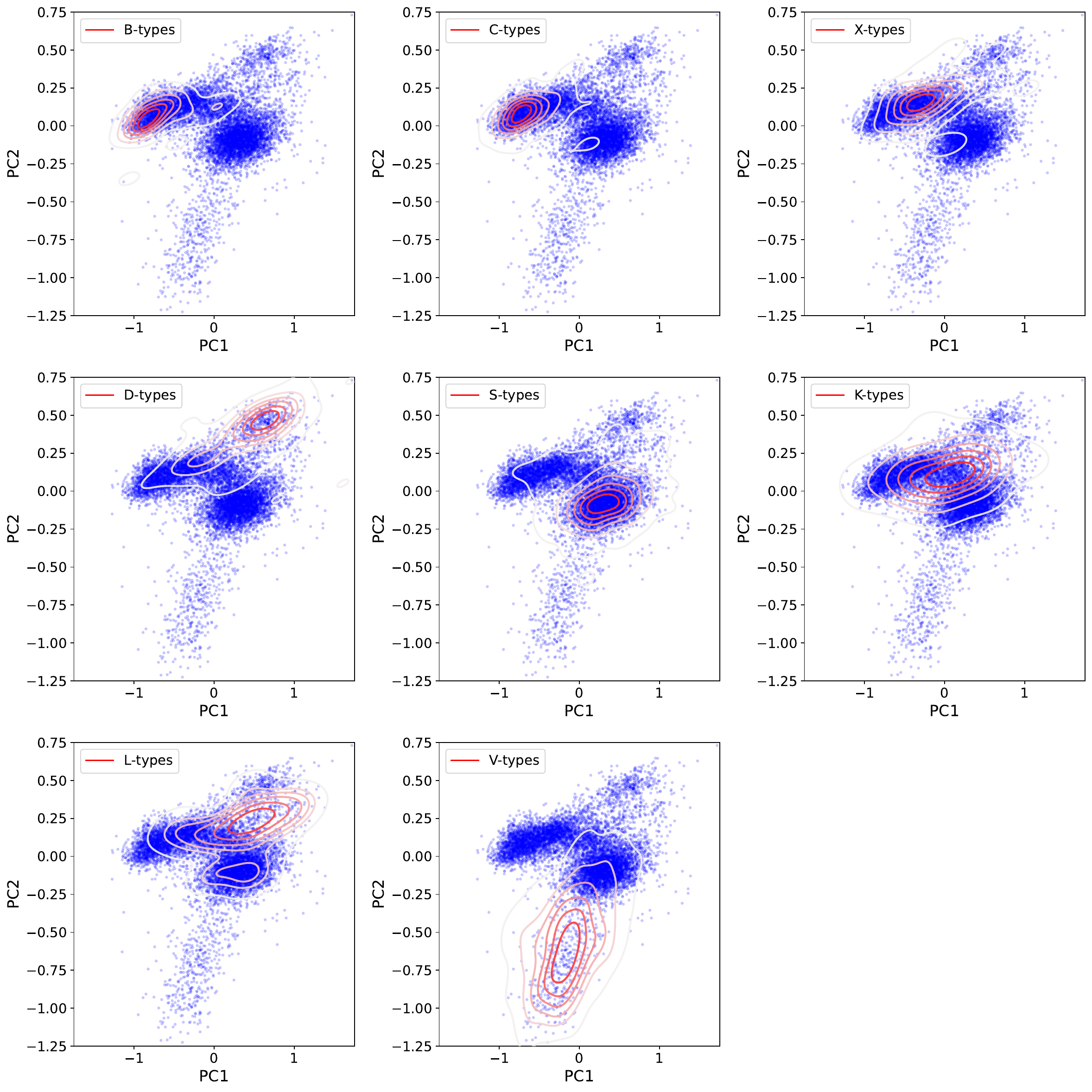}  
\caption{Distribution of several Bus-DeMeo taxonomic classes, as indicated in the plot legend, on the PC1--PC2 plane for asteroids with Gaia DR3 reflectance spectra and S/N $>$ 30 (shown as grey points).}  
\label{F:pcaClasses}
\end{figure*}

\end{appendix}
\end{document}